%% file: main.tex
\pdfoutput=1
\documentclass[conference]{IEEEtran}

\ifCLASSINFOpdf
\else
\fi

\input{includes_and_commands.tex}

\makeatletter

\begin{document}
\title{Spatial-Domain Wireless Jamming with Reconfigurable Intelligent Surfaces}

\newcommand{\linebreakand}{%
  \end{@IEEEauthorhalign}
  \hfill\mbox{}\par
  \mbox{}\hfill\begin{@IEEEauthorhalign}
}
\makeatother

\author{\IEEEauthorblockN{  Philipp Mackensen$^{1,}$\IEEEauthorrefmark{1}, 
                            Paul Staat$^{1,}$\IEEEauthorrefmark{2}, 
                            Stefan Roth\IEEEauthorrefmark{1}, 
                            Aydin Sezgin\IEEEauthorrefmark{1}, 
                            Christof Paar\IEEEauthorrefmark{2} and 
                            Veelasha Moonsamy\IEEEauthorrefmark{1}}
        \IEEEauthorblockA{\IEEEauthorrefmark{1}Ruhr University Bochum, Bochum, Germany\\
                          \IEEEauthorrefmark{2}Max Planck Institute for Security and Privacy, Bochum, Germany}
        \IEEEauthorblockA{E-Mail: \{philipp.mackensen, stefan.roth-k21, aydin.sezgin, veelasha.moonsamy\}@rub.de, \\ \{paul.staat, christof.paar\}@mpi-sp.org}
}

\IEEEoverridecommandlockouts
\makeatletter\def\@IEEEpubidpullup{6.5\baselineskip}\makeatother
\IEEEpubid{\parbox{\columnwidth}{
    Network and Distributed System Security (NDSS) Symposium 2025\\
    24 February - 28 February 2025, San Diego, CA, USA\\
    ISBN 979-8-9894372-8-3\\
    https://dx.doi.org/10.14722/ndss.2025.240440\\
    www.ndss-symposium.org
}
\hspace{\columnsep}\makebox[\columnwidth]{}}

\maketitle
\def\thefootnote{1}\footnotetext{These authors contributed equally to this work.}

\input{acronyms}

\input{sections/abstract}
\input{sections/introduction}
\input{sections/background}

\input{sections/overview}

\input{sections/evaluation}

\input{sections/discussion}

\input{sections/related_work}
\input{sections/conclusion}

\bibliographystyle{plain}
\bibliography{references}

\input{sections/appendix}

\end{document}

%% file: includes_and_commands.tex
\usepackage[table,xcdraw]{xcolor}

\usepackage{xspace}
\usepackage{tikz}
\usepackage{amsmath}

\usepackage{xcolor}
\def\BibTeX{{\rm B\kern-.05em{\sc i\kern-.025em b}\kern-.08em
    T\kern-.1667em\lower.7ex\hbox{E}\kern-.125emX}}

\usepackage[mode=buildnew]{standalone}
\usepackage{tikz}
\usetikzlibrary{patterns}
\usetikzlibrary{positioning,shapes.misc}
\usetikzlibrary{arrows,fit,calc,automata}

\usepackage{pifont}
\usepackage{placeins}
\usepackage{multirow}
\usepackage{comment}

\usepackage[nolist]{acronym}
\usepackage{amssymb}
\usepackage{amsfonts}
\usepackage{amsfonts}
\usepackage{amsthm}

\usepackage{amsmath}
\usepackage{float}
\usepackage{scrextend}

\usepackage{eurosym}
\usepackage{enumitem}

\usepackage[binary-units=true]{siunitx}

\usepackage{caption}
\usepackage{subcaption}
\usepackage{graphicx}

\usepackage{booktabs}

\usepackage{listings}

\usepackage{diagbox}
\usepackage{cancel}

\usepackage{colortbl}
\usepackage{zref-savepos}
\usepackage{pict2e}
\usepackage{wasysym}

\usepackage{array}
\usepackage{hhline}

\usepackage{xurl}
\usepackage{hyperref}

\newcommand{\cf}{\textit{cf}.\xspace}
\newcommand{\etal}[0]{\textit{et~al.}\xspace}
\newcommand{\eg}{\textit{e.g.}\xspace}
\newcommand{\ie}{\textit{i.e.}\xspace}

\newsavebox{\largestimage}

\newcounter{NoTableEntry}
\renewcommand*{\theNoTableEntry}{NTE-\the\value{NoTableEntry}}

\newlength{\plotwidth}
\makeatletter
\newcommand*{\notableentry}{%
  \kern-\tabcolsep
  \stepcounter{NoTableEntry}%
  \vadjust pre{\zsavepos{\theNoTableEntry t}}%
  \vadjust{\zsavepos{\theNoTableEntry b}}%
  \zsavepos{\theNoTableEntry l}%
  \raisebox{%
    \dimexpr\zposy{\theNoTableEntry b}sp
    -\zposy{\theNoTableEntry l}sp\relax
  }[0pt][0pt]{%
    \color{black}%
    \setlength{\unitlength}{1pt}%
    \edef\w{%
      \strip@pt\dimexpr\zposx{\theNoTableEntry r}sp%
      -\zposx{\theNoTableEntry l}sp\relax
    }%
    \edef\h{%
      \strip@pt\dimexpr\zposy{\theNoTableEntry t}sp%
      -\zposy{\theNoTableEntry b}sp\relax
    }%
    \ifdim\w pt=0pt
    \else
      \begin{picture}(0,0)%

        \edef\x{%
          \noexpand\put(0,0){\noexpand\line(\w,\h){\w}}%
          \noexpand\put(0,\h){\noexpand\line(\w,-\h){\w}}%
        }\x
      \end{picture}%
    \fi
  }%
  \hspace{0pt plus 1filll}%
  \zsavepos{\theNoTableEntry r}
  \kern-\tabcolsep
}

%% file: acronyms.tex
\begin{acronym}

\acro{AP}{access point}

\acro{CSI}{channel state information}

\acro{FDD}{frequency-division duplex}

\acro{RIS}{reconfigurable intelligent surface}

\acro{LoS}{line of sight}

\acro{MCS}{modulation and coding scheme}

\acro{OFDM}{orthogonal frequency division multiplexing}

\acro{PCB}{printed circuit board}

\acro{RF}{radio frequency}
\acro{RSSI}{received signal strength indicator}

\acro{SNR}{signal-to-noise ratio}
\acro{JSR}{jamming-to-signal ratio}
\acro{SJNR}{signal-to-jamming-and-noise ratio}

\acro{TDD}{time-division duplex}

\acro{WLAN}{wireless local area network}
\acro{WSN}{wireless sensor network}

\acro{VNA}{vector network analyzer}

\acro{V2X}{vehicle-to-everything}

\acro{MAC}{media access control}

\acro{MIMO}{multiple-input and multiple-output}

\acro{QoS}{quality of service}

\end{acronym}

%% file: sections/abstract.tex
\begin{abstract}

Wireless communication infrastructure is a cornerstone of modern digital society, yet it remains vulnerable to the persistent threat of wireless jamming. Attackers can easily create radio interference to overshadow legitimate signals, leading to denial of service. 
The broadcast nature of radio signal propagation makes such attacks possible in the first place, but at the same time poses a challenge for the attacker: The jamming signal does not only reach the victim device but also other neighboring devices, preventing precise attack targeting.

In this work, we solve this challenge by leveraging the emerging \acf{RIS} technology, for the first time, for precise delivery of jamming signals. In particular, we propose a novel approach that allows for environment-adaptive spatial control of wireless jamming signals, granting a new degree of freedom to perform jamming attacks.  
We explore this novel method with extensive experimentation and demonstrate that our approach can disable the wireless communication of one or multiple victim devices while leaving neighboring devices unaffected. Notably, our method extends to challenging scenarios where wireless devices are very close to each other: We demonstrate complete denial-of-service of a \mbox{Wi-Fi} device while a second device located at a distance as close as \SI{5}{\mm} remains unaffected, sustaining wireless communication at a data rate of \SI{25}{Mbit/s}. Lastly, we conclude by proposing potential countermeasures to thwart RIS-based spatial domain wireless jamming attacks.

\end{abstract}

%% file: sections/introduction.tex
\section{Introduction}
\label{sec:introduction}

Wireless communication systems are ubiquitous and seamlessly provide connectivity to the smart and interconnected devices that permanently surround us. In our modern daily lives, we frequently use instant messaging, media streaming, health monitoring, and home automation -- all of which rely on wireless systems and their constant availability. 
However, wireless systems utilize a broadcast medium that is open to everyone, inherently exposing a large attack surface. One particular critical threat is \emph{wireless jamming}, which allows malicious actors to perform denial of service attacks with minimal effort. In a classical jamming attack, the adversary transmits an interfering signal that overshadows the desired signal, preventing a victim receiver from correctly decoding it. Crucially, loss of connectivity impacts the functionality of wireless devices and can thus have potentially far-reaching consequences, such as in smart grids, smart transportation, and healthcare systems. Recent media reports underscore the real-world threat potential of jamming attacks, \eg, criminals disabling smart home security systems~\cite{smart_home_thief_1, smart_home_thief_2} and preventing cars from locking~\cite{car_jamming}.

This basic attack principle has previously been studied by a large body of research: For instance, the attacker can leverage various jamming waveforms, such as noise or replayed victim signals~\cite{hangPerformanceDSSSRepeater2006}, and vary the attack timing, jamming constantly~\cite{xuFeasibilityLaunchingDetecting2005} or only at certain times~\cite{schulzMassiveReactiveSmartphoneBased2017}. As evident from the many existing attack strategies~\cite{lichtmanCommunicationsJammingTaxonomy2016, xuFeasibilityLaunchingDetecting2005, poiselModernCommunicationsJamming2011, pelechrinisDenialServiceAttacks2010}, wireless jamming has been incrementally refined and became increasingly sophisticated. One particular example for this is the case of selective jamming attacks.

To illustrate a potential attack scenario, consider an adversary attempting to sabotage a complex automated manufacturing process. Distributed actuators might take orders from several previous processing stages that have to be executed in a timely fashion, risking manufacturing failure otherwise. Here, the adversary could use selective jamming to simulate local loss of connectivity on a single actuator but not the entire plant which would likely trigger some emergency response.

So far, the only means to realize such a selective jamming attack is via so-called reactive jamming. Here, the attacker analyzes all wireless traffic in real-time to decide on-the-fly whether to send a jamming signal~\cite{schulzMassiveReactiveSmartphoneBased2017, proanoSelectiveJammingAttacks2010, arasSelectiveJammingLoRaWAN2017}, relying on the existence of meaningful protocol-level information not protected by cryptographic primitives. %
In our manufacturing plant example, selective disruption of the actuator would require the attacker to receive and identify \emph{every} packet directed to the recipient before sending a jamming signal. This restricts the attacker positioning rather close to the victim. Other downsides of this approach are that it can be mitigated by fully disguising packet destinations and the attack realization being rather complex and cumbersome.

In light of these aspects, we are interested in novel attack strategies resolving the aforementioned shortcomings. Clearly, the ideal solution would be to physically inject a proactive jamming signal directly and only into the victim device which, however, is not possible due to the wireless nature of jamming and the inevitable broadcast behavior of radio signal propagation to other, non-target devices. Thus, we aim to answer the following research question:%
\begin{enumerate}[start=1,label={}, leftmargin = 1em, series=rqc]
    \item \emph{How can we physically target and jam one device while keeping others operational?}
\end{enumerate}

We solve this challenge by means of a \acf{RIS} to devise the first selective jamming mechanism based on taming random wireless radio wave propagation effects. Using \ac{RIS}-based environment-adaptive wireless channel control, allowing to maximize and minimize wireless signals on specific locations~\cite{kainaShapingComplexMicrowave2014}, the attacker gains spatial control over their wireless jamming signals. This opens the door to precise jamming signal delivery towards a target device, disrupting any legitimate signal reception, while leaving other, non-target devices, untouched. Other than reactive jamming, this is a true physical-layer selection mechanism, allowing realization independent of protocol-level information. Moreover, the attacker only initially needs to detect signals from considered devices, removing the need for any real-time monitoring and reaction to ongoing transmissions.

In this work, we experimentally evaluate \ac{RIS}-based spatially selective jamming attacks against \mbox{Wi-Fi} communication, showing that it is possible to target one or multiple devices while keeping non-target devices operational. To accomplish this, we exploit that considered devices transmit signals, allowing the attacker to passively adapt to the scene.
Apart from the attack's core mechanism, we study crucial real-world aspects such as the attack's robustness against environmental factors.
We additionally verify the effectiveness of our attack in real-world wireless networks, where mechanisms that could counteract the attack are at play, \eg, adaptive rate control of \mbox{Wi-Fi} networks.
We show that \ac{RIS}-based selective jamming even works despite extreme proximity of devices, \eg, \SI{5}{mm}, and investigate the underlying physical mechanisms. Finally, we perform comparison experiments with a directional antenna, showing significant of our \ac{RIS}-based approach.

In summary, our work makes the following key contributions:
\begin{itemize}[nosep]
    \item We propose the first true physical-layer selective targeting mechanism for wireless jamming, enabling environment-adaptive attacks in the spatial domain.

    \item We present an attack realization based on \ac{RIS}, using passive eavesdropping to determine an appropriate \ac{RIS} configuration which is the key to deliver jamming jamming signals towards targeted devices while avoiding non-target devices.

    \item We present a comprehensive experimental evaluation with commodity \mbox{Wi-Fi} devices, environmental changes, and an in-depth analysis of the physical properties of our jamming attack.

\end{itemize}

%% file: sections/background.tex
\section{Technical Background}
\label{sec:background}

In this section, we introduce the necessary background on wireless jamming and \acp{RIS}.

\subsection{Wireless Jamming}
\label{sec:background:jamming}

Wireless communication quality of service, \eg, reliability and data throughput, is determined by the quality of received signals. An example is weak signal reception which yields a low \ac{SNR} and consequently, increases the probability of bit errors. Similarly, when multiple radios transmit simultaneously, the receiver observes the superposition of multiple signals, \ie, the desired signal with additional interference, again degrading performance. A \textit{wireless jamming} attacker exploits this mechanism by deliberately sending strong interfering signals. At the victim receiver, they overshadow legitimate signals, increase the bit error probability, and eventually lead to complete denial of service~\cite{poiselModernCommunicationsJamming2011, groverJammingAntijammingTechniques2014, lichtmanCommunicationsJammingTaxonomy2016}.

There exist several types of jamming attacks, differing in terms of, \eg, specifics of the attack target, the used interfering signal waveform, or the level of cognition~\cite{groverJammingAntijammingTechniques2014}. The jamming signal can comprise of, \eg, noise, single or multi-tone carriers, or valid waveforms carrying digital information. While noise jamming reduces the \ac{SNR}, valid signals can enhance attack effectiveness, \eg, by keeping the receiver busy~\cite{gvozdenovicTruncatePreamblePHYBased2020}. Attackers may constantly or reactively transmit the jamming signal, \ie, upon detecting a certain trigger~\cite{girkeResilient5GLessons2019}. %

\subsection{Reconfigurable Intelligent Surfaces}
\label{sec:background:irs}

\begin{figure}
    \centering
    \includegraphics[width=0.9\columnwidth]{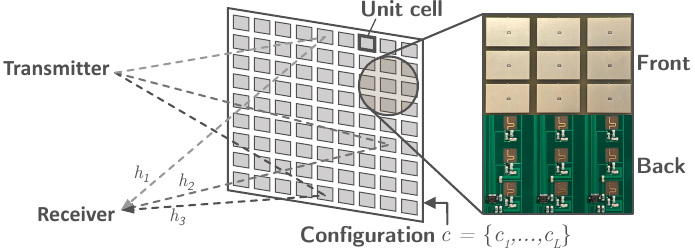}
    \caption{Illustration of the \ac{RIS} operation principle along with photos of the~\ac{RIS} hardware implementation~\cite{heinrichsOpenSourceReconfigurable2023}, where the configuration vector $c$ determines the radiation behavior.}
    \label{fig:background:ris:reflectarray}
\end{figure}

An \ac{RIS} is an engineered surface to digitally control reflections of radio waves, enabling \textit{smart radio environments}. It is worth noting that \acp{RIS} are likely to become pervasive, as they hold the potential to complement future wireless networks such as 6G~\cite{jiangRoad6GComprehensive2021, chowdhury6GWirelessCommunication2020, tataria6GWirelessSystems2021}. Here, the propagation medium is considered as a degree of freedom to optimize wireless communication by redirecting radio waves in certain directions~\cite{liaskosNovelCommunicationParadigm2019}, \eg, to improve signal coverage and eliminate dead zones, to enhance energy efficiency and data throughput~\cite{liu2021reconfigurable}, and building low-complexity base stations~\cite{basarWirelessCommunicationsReconfigurable2019}.

An \ac{RIS} does not actively generate its own signals but passively reflects existing ambient signals. For this, it utilizes $L$~identical unit-cell reflector elements arranged on a planar surface, as shown in Figure~\ref{fig:background:ris:reflectarray}. Importantly, the reflection coefficient of each reflector is separately tunable to shift the reflection phase. Typically, an \ac{RIS} is realized as a \ac{PCB} with printed microstrip reflectors, enabling very low-cost implementation. To reduce complexity, many \acp{RIS} use \SI{1}{\bit} control~\cite{ranaReviewPaperHardware2023}, \ie, to select between two reflection phases \SI{0}{\degree} and \SI{180}{\degree}, corresponding to the reflection coefficients~$+1$ and~$-1$. This allows the control circuitry to directly interface with digital logic signals from, \eg, a microcontroller. %
The technology is still under development~\cite{ranaReviewPaperHardware2023} which is why \acp{RIS} are currently not widely used in practice. At the time of writing, first implementations are being made commercially available~\cite{greenerwave_website, novoflect_website} and field trials are being carried out~\cite{peiRISaidedWirelessCommunications2021}.

\subsubsection{Finding \ac{RIS} configurations} 
\label{sec:background:irs:optimization}

To realize a desired \ac{RIS} reflection behavior between a sender and a receiver, an appropriate \ac{RIS} configuration is required that matches the radio environment. For example, to maximize signal power at a receiver, the \ac{RIS} configuration is used to make all signal components traveling via the \ac{RIS} add coherently with other non-\ac{RIS} signal components.
Usually, such an \ac{RIS} configuration cannot be be blindly synthesized due to the complexity of scene-dependent and hard-to-predict radio wave propagation effects in conjunction with the vast configuration space of the \ac{RIS}. That is, an $L$-element binary-tunable \ac{RIS} has $2^L$ possible configurations. Therefore, \ac{RIS} configurations are often determined based on iterative optimization algorithms, involving measurement feedback to assess how a particular \ac{RIS} configuration influences the wireless channel~\cite{feng2021optimization, zou2021robust, tewes2022full, kainaShapingComplexMicrowave2014, peiRISaidedWirelessCommunications2021}.

%% file: sections/overview.tex
\section{Preliminaries}
\label{sec:overview}

\subsection{Threat Model}
\label{sec:overview:threat_model}

\subsubsection{Attack Scenario}
We consider a typical wireless network scenario where a number of wireless devices are deployed and connected to an \ac{AP}. At least temporarily, the devices are stationary and do not change location. 
We assume the devices communicate with the \ac{AP} using \mbox{Wi-Fi}, but the following analysis holds for any \ac{TDD} communication protocol. Additionally, we assume reciprocity of wireless channels, meaning that for a pair of devices, the same radio propagation effects  occur, regardless of the communication direction. Finally, we assume that the wireless signals are subject to multipath propagation due to typical propagation phenomena, \eg, reflection and scattering, as commonly found in indoor and urban environments. %

\subsubsection{Attacker Model}
We consider a physical-layer wireless jamming attacker who generates radio interference with the goal of disrupting the wireless communication of a set of victim devices. The attacker aims to perform selective jamming, meaning they aim to disrupt only a subset of devices while leaving others unaffected. 

The attacker is capable of transmitting and receiving radio signals towards and from considered devices, \eg, by using an ordinary radio transceiver comparable to the hardware of the considered devices. We assume that the attacker utilizes a single antenna to either transmit or receive signals. Additionally, the attacker employs an \ac{RIS} next to their antenna, which they can configure arbitrarily. %

The attacker is external to the wireless network of the considered devices and cannot read encrypted payload information. However, the attacker can estimate the \ac{RSSI} and distinguish signals originating from different devices. Finally, the attacker can choose an arbitrary position to launch their attack without knowing the exact location of the considered devices.

\subsection{System Model}
\label{sec:system_model}
In this section, we establish the system model to formally describe the attack scenario and the involved parties from a signal perspective. %
For the reader's convenience, we summarize the used symbols in~\autoref{tab:symbol_table}.

\begin{table}
\centering
\caption{Terminology Overview}
\label{tab:symbol_table}
\scriptsize 
\begin{tabular}{|l|l|}
\hline
\textbf{Symbol} & \textbf{Description} \\ \hline \hline
$\mathcal{D}$ & Set of all devices  \\ \hline
$D_i$ & $i^{th}$ device in $\mathcal{D}$ \\ \hline
$N$ & Number of total devices in $\mathcal{D}$ \\ \hline
$\mathcal{T}$ & Subset of target devices \\ \hline
$\mathcal{N}$ & Subset of non-target devices \\ \hline
$L$ & Number of IRS elements \\ \hline
$c$ & \ac{RIS} configuration vector\\ \hline
$c_l$ & Reflection coefficient of the $l^{th}$ \ac{RIS} element \\ \hline
$H^{T}_{R}$ & Channel gain from sender $T$ to receiver $R$ \\ \hline
$H^{T}_{R}(c)$ & \ac{RIS}-controlled channel gain \\ \hline 
$h_{l}^{D_i}$ & $l^{th}$ RIS sub-channel to device $D_i$\\ \hline
$X_T$ & Signal from a sender $T$\\ \hline
$J$ & Jamming signal from the attacker \\ \hline
$W$ & White Gaussian noise \\ \hline
\end{tabular}
\end{table}

\begin{figure}
    \centering
    \includegraphics[width=0.8\plotwidth]{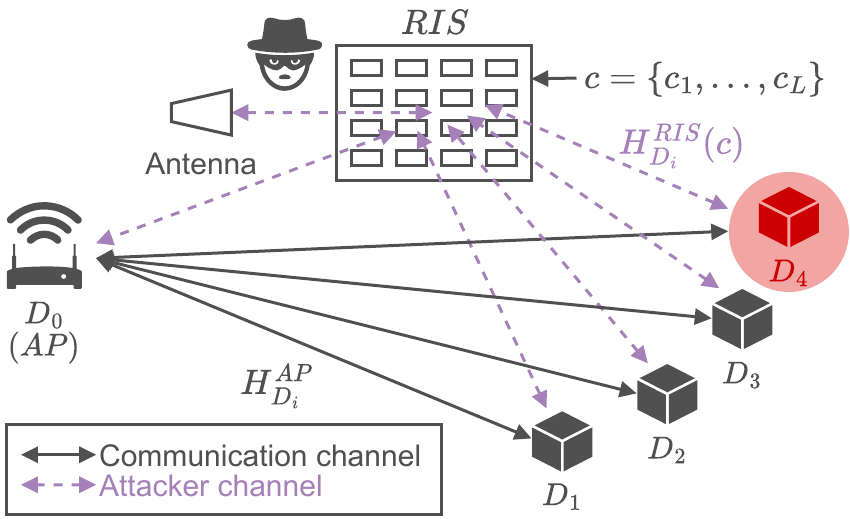}
    \caption{Illustration of the assumed system model.}
    \label{fig:system_model}
\end{figure}

We assume that the attacker faces a total of $N$ wireless devices from the set $\mathcal{D} = \{D_0, \dots, D_{N-1}\}$, \eg, forming a wireless network where one device is an \ac{AP} that the others connect to. The devices $\{D_1, \dots, D_{N-1}\}$ seek to extract correct digital information from the legitimate signal~$X_{AP}$ received from the \ac{AP}~$D_0$. The attacker seeks to disrupt the wireless communication of $K$ devices, forming the subset $\mathcal{T} \subseteq \mathcal{D}$, while leaving the remaining %
devices in the subset $\mathcal{N} = \mathcal{D} \setminus \mathcal{T}$ unaffected. Figure~\ref{fig:system_model} illustrates an exemplary scenario with five devices $\mathcal{D} = \{D_0,\ D_1,\ D_2,\ D_3,\ D_4\}$, where the attacker would like to jam $\mathcal{T} = \{D_4\}$, while keeping the remaining devices $\mathcal{N} = \{D_0,\ D_1,\ D_2,\ D_3\}$ operational.

To achieve their goal, the attacker transmits a jamming signal $J$ to overshadow the legitimate signal from the \ac{AP}~$X_{AP}$. 
Both signals $X_{AP}$ and $J$ are subject to radio propagation effects, described by the complex channel gains between the respective transmitter and the device $D_i$, denoted as $H^{AP}_{D_i}$ from the \ac{AP} to $D_i$, and $H^{RIS}_{D_i}(c)$ from the attacker to $D_i$. Thus, the device $D_i$ observes the total received signal
\begin{align}
    Y_{D_i} &= H^{AP}_{D_i} X_{AP} + H^{RIS}_{D_i}(c)\ J + W,
    \label{eq:di_rx_signal}
\end{align}
where $W$ is additive white Gaussian noise. Note that the attacker's channel is reconfigurable by means of the \ac{RIS} configuration vector $c$. In line with the literature~\cite{basarWirelessCommunicationsReconfigurable2019}, we model this channel as the superposition of $L$ sub-channels $h^{D_i}_l$ between the attacker's antenna and the device $D_i$ via the $l^{th}$~\ac{RIS} element, each of which is multiplied with the selected reflection coefficient $c_l$ of the respective \ac{RIS} reflector element:
\begin{equation}
    H^{RIS}_{D_i}(c) = \sum_{l=1}^{L} h^{D_i}_l\, c_l.
\label{eq:attacker_channel}
\end{equation}

From \autoref{eq:di_rx_signal}, we can formulate the \ac{JSR} of each device as
\begin{equation}
    \mathrm{JSR_{D_i}} = \frac{|H^{RIS}_{D_i}(c) J|^2}{|H^{AP}_{D_i} X_{AP}|^2},
    \label{eq:jsr}
\end{equation}
which is a key metric to assess the success of jamming attacks~\cite{poiselModernCommunicationsJamming2011, pelechrinisDenialServiceAttacks2010}. With increasing \ac{JSR}, the probability that the respective radio receiver will be disturbed increases.

\section{\ac{RIS}-based Selective Jamming Attack Strategy}
\label{sec:attack_strategy}

With the established system model in mind, we now proceed to elaborate the attacker's strategy in order to meet their two principal goals: ($i$)~rendering target devices inoperative while ($ii$)~keeping non-target devices operational. For this, the attacker must maximize $\mathrm{JSR_{D_i}}$ for the target devices and minimize it for the non-target devices. The classical approach to meet goal~($i$) is to increase the power of the jamming signal~$J$. However, this strategy does not address goal~($ii$) and carries the risk of also jamming non-target devices. In this work, we resolve this issue by means of an \ac{RIS}, leveraging for the first time \ac{RIS}-based wireless channel control to optimize an active jamming attack. In particular, the attacker leverages the \ac{RIS} configuration~$c$ to adapt their wireless channel gains $H^{RIS}_{D_i}(c)$ and control the delivery of $J$ towards each device. 
In other words, by applying an appropriate configuration~$c$ to their \ac{RIS}, the attacker can selectively increase or decrease the channel gains towards the considered devices in order to control the respective \ac{JSR} and thereby control the effect of the jamming. Thus, the attacker faces the following multivariate optimization problem:
\begin{align}
    \max_{c}\  &|H^{RIS}_{d \in \mathcal{T}}(c)|,\label{eq:maxH}\\
    \min_{c}\  &|H^{RIS}_{d \in \mathcal{N}}(c)|.\label{eq:minH}
\end{align}

To find an appropriate $c$ that meets these goals, the attacker must observe $H^{D_i}_{RIS}(c)$~(see~\autoref{sec:background:irs}) -- ideally by measuring the jamming signal strength arriving at each considered device. However, this clearly is not possible as the devices $\mathcal{D}$ do not cooperate with the attacker. To solve this, we leverage \textit{channel reciprocity}, where the wireless channels from the attacker to the considered devices and vice versa are identical, \ie, it holds that $H^{D_i}_{RIS}(c) \approx H^{RIS}_{D_i}(c)$~\cite{tang2021channel}. Consequently, to assess whether a particular \ac{RIS} configuration $c$ meets the channel optimization, the attacker can eavesdrop on the considered devices and measure $H^{D_i}_{RIS}(c)$.

\begin{figure}
\centering
    \begin{subfigure}{.5\columnwidth}
        \centering
        \includegraphics[width=\columnwidth]{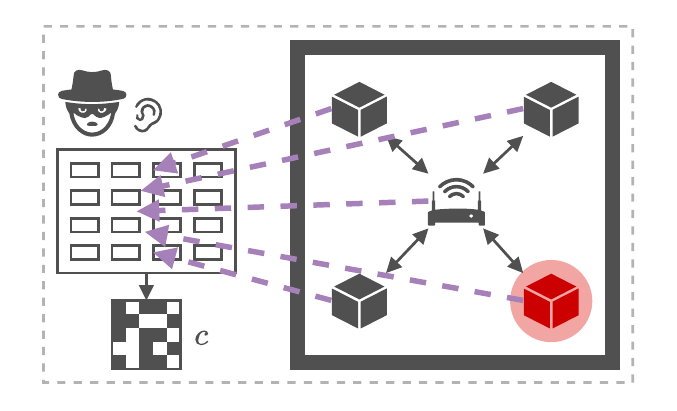}
        \caption{}
        \label{fig:two_step_attack:step_1}
    \end{subfigure}%
    \begin{subfigure}{.5\columnwidth}
        \centering
        \includegraphics[width=\columnwidth]{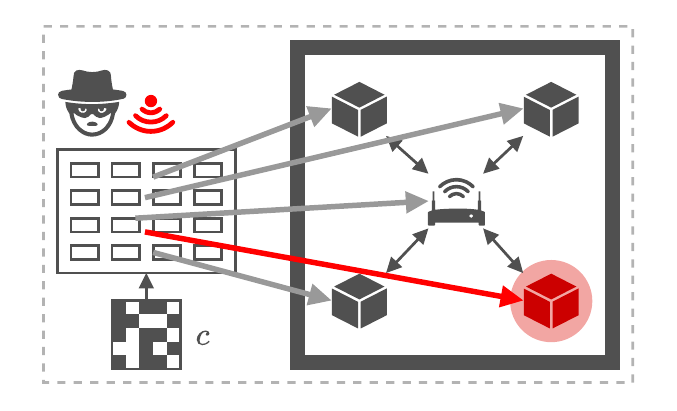}
        \caption{}
        \label{fig:two_step_attack:step_2}
    \end{subfigure}
    \caption{Two-step attack strategy of the jamming attack using the \ac{RIS}. (a) Step 1: Passive channel optimization. (b) Step 2: Active wireless jamming attack.} 
    \label{fig:two_step_attack}
\end{figure}

In summary, the \ac{RIS}-based jamming attack is a two-step procedure as illustrated in Figure~\ref{fig:two_step_attack}. First, the attacker passively determines an appropriate \ac{RIS} configuration by eavesdropping the radio communication signals from the considered devices~(see Figure~\ref{fig:two_step_attack:step_1}). Using this configuration, the attacker then actively transmits the jamming signal $J$ which disrupts target devices while non-target devices remain operational~(see Figure~\ref{fig:two_step_attack:step_2}).

%% file: sections/evaluation.tex
\section{Experimental Setup}
\label{sec:overview:experimental_setup}

\subsubsection{Wireless Environment}

\begin{figure}%
    \centering
    \includegraphics[width=0.70\plotwidth]{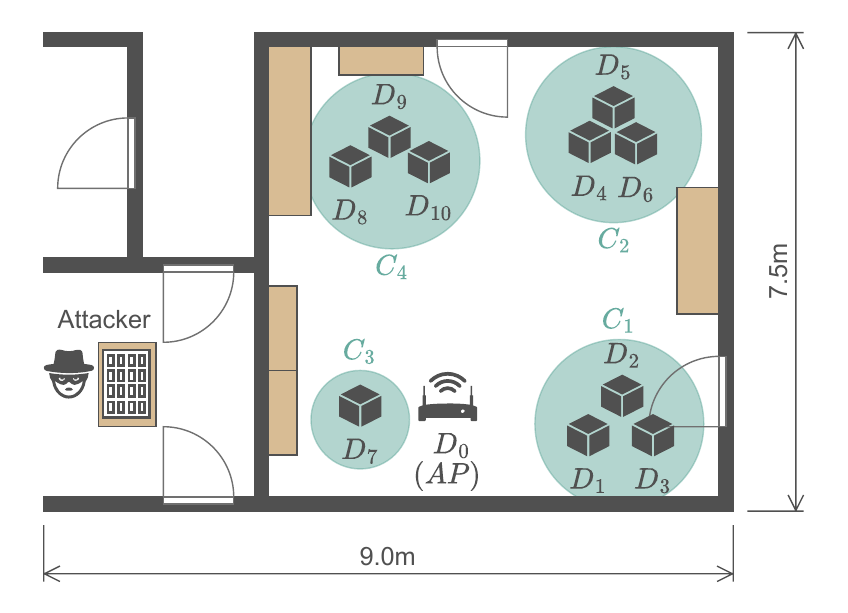}
    \caption{Floorplan of the environment used for experiments, indicating the positions of all wireless devices and the attacker.}
    \label{fig:experimental_setup}
\end{figure}

We conduct our experiments in an ordinary office environment where we make use of an area of approximate size \SI{9.0}{m}~$\times$~\SI{7.5}{m}. A floor plan is depicted in Figure~\ref{fig:experimental_setup}, indicating the position of the attacker and the wireless devices $\{D_0, \dots, D_{10}\}$. The devices are arranged in four clusters spread across the room.%

To demonstrate and evaluate the attack, we utilize commodity off-the-shelf \mbox{Wi-Fi} devices. For the \ac{AP} device $D_0$, we utilize a TP-Link~N750 router running OpenWrt to provide an IEEE~802.11n \mbox{Wi-Fi} network with \SI{20}{\MHz} bandwidth on channel~$112$, corresponding to a channel center frequency of~\SI{5560}{\MHz}. The router is capable of packet injection using lorcon~\cite{lorcon}. For the wireless devices $\{D_1, \dots, D_{10}\}$, we use ten \mbox{Raspberry Pi 4 Model B} and leverage nexmon~\cite{nexmon:project} to optionally put their \mbox{Wi-Fi} chipset into monitor mode. 

The devices $D_1$ to $D_{10}$ ping the \ac{AP} ($D_0$) to trigger wireless traffic. Per default, the \ac{AP} is part of the set of non-target devices $\mathcal{N}$. %

\subsubsection{Attacker Setup}
On the attacker side, we use the following hardware setup. To realize eavesdropping, we employ a \mbox{Raspberry~Pi~4 Model B}. Again, we use nexmon~\cite{nexmon:project} to put the \mbox{Wi-Fi} chipset into monitor mode, allowing us to obtain the \ac{RSSI} value and MAC~address for each received \mbox{Wi-Fi} packet, even in the case of frame errors. In order to connect an external antenna to the Raspberry Pi, we disconnected the onboard \ac{PCB} antenna and added a coaxial connector.   %

For the active jamming, we utilize standard IEEE~802.11n \mbox{Wi-Fi} signals (\SI{20}{\MHz} bandwidth with \ac{MCS} set to~1), containing~$25$ randomized payload bytes. For convenient signal generation, we use a Signal~Hound~VSG60 vector signal generator, allowing to transmit signals with a \SI{100}{\percent} duty cycle and precisely controlled signal power. However, we stress that jamming signal generation can likewise be realized with ordinary \mbox{Wi-Fi} devices~\cite{schulzMassiveReactiveSmartphoneBased2017}. %

We employ an~interline~PANEL~14~directional antenna and use a Mini-Circuits~USB-2SP4T-63H \ac{RF} switch to either connect the antenna to the Raspberry Pi for eavesdropping or to the signal generator for jamming.
The antenna is directed towards the attacker's~\ac{RIS}. \autoref{fig:jamming_setup:photos:attacker} in the~Appendix~\ref{appendix:experimental_setup} shows a photo of the setup. The \ac{RIS} is based on the open-source design of Heinrichs~\etal~\cite{heinrichsOpenSourceReconfigurable2023} and consists of three standard FR4 \acp{PCB}. It has $L=768$ unit-cell reflector elements with binary phase control, optimized to operate in the \SI{5}{\GHz} \mbox{Wi-Fi} frequency range. The elements can be programmed via USB to select the phase of $c_l$ to either be \SI{0}{\degree} (state~`\texttt{0}') or \SI{180}{\degree} (state~`\texttt{1}'). For further technical details, we refer to~\cite{heinrichsOpenSourceReconfigurable2023}.

\subsubsection{\ac{RIS} Optimization}
\label{sec:overview:experimental_setup:ris_optimization}
To determine an \ac{RIS} configuration that solves the optimization problem formulated in \autoref{eq:maxH} and~\autoref{eq:minH}, we employ the greedy genetic optimization algorithm put forward by Tewes~\etal~\cite{tewes2022full}. The algorithm stores a sorted table, where a cost function $f$ is evaluated for a set of $B$ initially random \ac{RIS} configurations. The cost function first aggregates the \ac{RSSI} values from the devices in the respective set, using weighted combinations of the mean and minimum for $\mathcal{T}$, and mean and maximum for $\mathcal{N}$. We weight the mean with $0.3$ and the extreme values with $0.7$, to emphasize the worst-performing devices in the respective sets stronger during optimization. Finally, we take the signed squared difference of both aggregate values as the result of~$f$. Based on \ac{RIS}-element-wise empirical probabilities for maximizing the cost function within the table, a new \ac{RIS} configuration is generated and evaluated to update the table with every algorithm step. For our experiments, we set $B$ to $100$ and run the algorithm for \SI{10000}{}~steps.

\subsubsection{Evaluation Metrics}

In the remainder of the paper, we use the following evaluation metrics:

\begin{itemize}
    \item \emph{\ac{RSSI} values}: The Raspberry~Pis we use provide estimates of the received signal strength in dBm with a resolution of \SI{1}{dB} for every received packet. We use \ac{RSSI} for channel measurement, \eg, to estimate $|H^{D_i}_{RIS}(c)|$ as needed to optimize the attacker's \ac{RIS}.
    
    \item \emph{\ac{JSR}}: We evaluate $\mathrm{JSR_{D_i}}$, \cf~\autoref{eq:jsr} for each device, using \ac{RSSI} values corresponding to signals received by the devices $\{D_1, \dots, D_{10}\}$ from the attacker and the \ac{AP}.

    \item \emph{Packet rates}: For evaluation of the attacker's selective jamming capabilities, we leverage packet injection on the \ac{AP} $D_0$ to transmit \mbox{Wi-Fi} packets with \ac{MCS}~6 at a constant rate of $100$~packets per second. On the devices $\{D_1, \dots, D_{10}\}$, we measure the successfully received \mbox{Wi-Fi} packets from the \ac{AP} per second. If a particular device is affected by the attacker's jamming signal $J$, the received packet rate will be reduced. For the measurements, the devices operate in monitor mode, granting a clear view on the bare jamming effects, independent of adaptive \mbox{Wi-Fi} mechanisms such as rate control, re-transmissions, or even disconnections.

    \item \emph{Data throughput}: We also evaluate the effect of the selective jamming in a standard \mbox{Wi-Fi} network, \ie, where devices do not use monitor mode but operate as a station. Here, $\{D_1, \dots, D_{10}\}$ are connected to the \ac{AP} $D_0$ and we assess the effect of the attacker's jamming signal $J$ by measuring the data throughput from the \ac{AP} towards a particular device in Mbit/s by means of \emph{iperf3}~\cite{iperf}.

\end{itemize}

\section{Attack Evaluation}
\label{sec:attack_evaluation}

After introducing the attack strategy and our experimental setup, we evaluate \ac{RIS}-based spatially selective jamming attacks in several real-world scenarios. First, we investigate selective jamming of a single device to demonstrate the scheme's feasibility. Then, we target multiple devices to show its scalability. We also assess the robustness of the \ac{RIS} optimization against environmental changes. In addition, we validate the attack's effectiveness in a fully-fledged Wi-Fi network and study devices in extreme proximity. Finally, we compare the performance of spatially selective jamming using a directional antenna versus the \ac{RIS}.

\subsection{Single-Target Jamming}
\label{sec:attack_evaluation:selective_jamming}

The first scenario we evaluate is jamming of a single target device. For this, we first optimize the attacker's \ac{RIS} and then transmit the jamming signal while evaluating how it affects each considered device $\{D_1, \dots, D_{10}\}$. Serving as a blueprint for the subsequent scenarios, we provide a detailed outline of the attacker's action, covering \ac{RIS} optimization, active jamming and \ac{JSR} analysis.

\subsubsection{\ac{RIS} Optimization and Active Jamming}

The attacker's first step is to find an appropriate configuration for the \ac{RIS} to maximize the received \ac{RSSI} from the targeted device while minimizing it for all other devices. For this, the attacker uses their Raspberry Pi to eavesdrop on the signals transmitted by the considered devices and estimates the magnitude of the channels $H^{D_i}_{RIS}(c)$ from the obtained \ac{RSSI} values. Once \ac{RSSI} values from all devices have been opportunistically collected, the attacker can perform a step of the \ac{RIS} optimization algorithm. In our experiments, the algorithm runtime for \SI{10000}{}~steps is about~\SI{5}{}~minutes.%

\begin{figure}
    \begin{subfigure}{.49\columnwidth}
        \centering
        \includegraphics[width=\columnwidth]{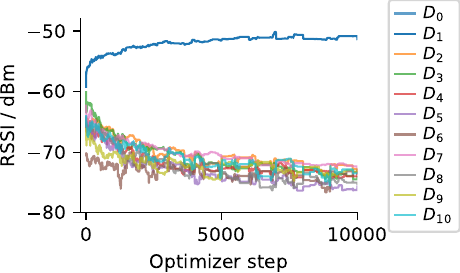}
        \caption{}
        \label{fig:evaluation:selective_jamming:single_target:optimization}
    \end{subfigure}%
    \begin{subfigure}{.49\columnwidth}
        \centering
        \includegraphics[width=\columnwidth]{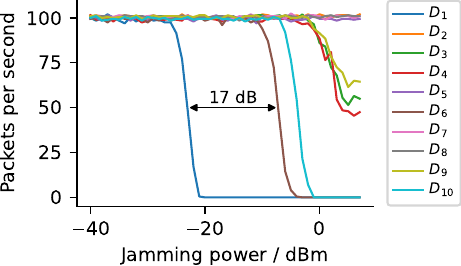}
        \caption{}
        \label{fig:evaluation:selective_jamming:single_target:jamming}
    \end{subfigure}
    \caption{(a) \ac{RIS} optimization process for targeting device $D_1$ while excluding all other devices. (b) Measurement of packet reception rates on the considered devices over jamming signal power using the previously optimized \ac{RIS} configuration.}
    \label{fig:evaluation:selective_jamming:single_target}
\end{figure}

In our initial experiment, we want to jam device $D_1$ and thus first need to find an \ac{RIS} configuration that targets device $D_1$ while excluding the remaining devices. \autoref{fig:evaluation:selective_jamming:single_target:optimization} depicts the \ac{RIS} optimization process, clearly showing an improvement in the channel quality between the \ac{RIS} and the target device $D_1$ while it degrades for the remaining devices as the optimization algorithm progresses. Eventually, the \ac{RSSI} for $D_1$ converges to \SI{-50}{dBm}, while the \ac{RSSI} for the non-targeted devices reach levels around \SI{-75}{dBm}, confirming the effectiveness of the optimization algorithm.

Using the \ac{RIS} configuration resulting from the optimization algorithm, the attacker switches from eavesdropping to actively sending a jamming signal. \autoref{fig:evaluation:selective_jamming:single_target:jamming} shows the packet reception rates from the \ac{AP} for each device over the jamming signal power. Here, we can see that jamming with a signal power of \SI{-21}{dBm} completely disrupts the reception of $D_1$, while the reception rates of all other devices are unaffected. Moreover, the attacker has a jamming signal power margin of \SI{17}{dB} until any other device ($D_6$) is disrupted.

\subsubsection{Jamming Success of the Attacker}
\label{sec:evaluation:jsr}
Thus far, we have demonstrated jamming of $D_1$ without affecting the other devices. However, the attacker can likewise leverage the \ac{RIS} optimization to target any other device in the environment. We repeat the previously outlined optimization process for the remaining devices $\{D_2, \dots, D_{10}\}$ and investigate the overall attack success by studying the \ac{JSR}.

\begin{figure}
\centering
    \begin{subfigure}{0.49\columnwidth}
        \centering
        \includegraphics[width=\columnwidth]{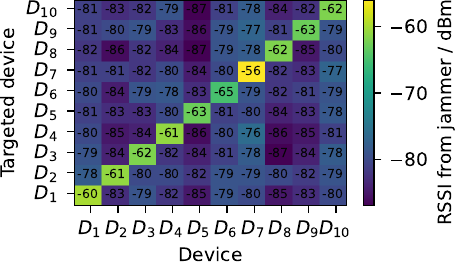}
        \caption{}
        \label{fig:evaluation:selective_jamming:optimization:rssi_attacker}
    \end{subfigure}
    \begin{subfigure}{0.49\columnwidth}
        \centering
        \includegraphics[width=\columnwidth]{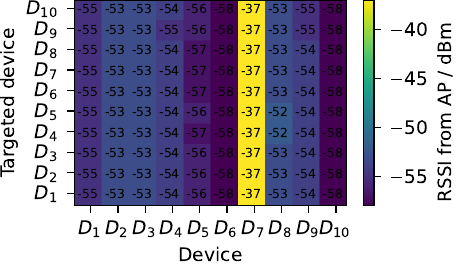}
        \caption{}
        \label{fig:evaluation:selective_jamming:optimization:rssi_accesspoint}
    \end{subfigure}\\
    
    \begin{subfigure}{0.49\columnwidth}
        \centering
        \includegraphics[width=\columnwidth]{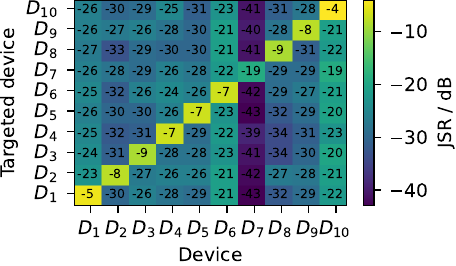}
        \caption{}
        \label{fig:evaluation:selective_jamming:optimization:jsr_raw}
    \end{subfigure}
    \begin{subfigure}{0.49\columnwidth}
        \centering
        \includegraphics[width=\columnwidth]{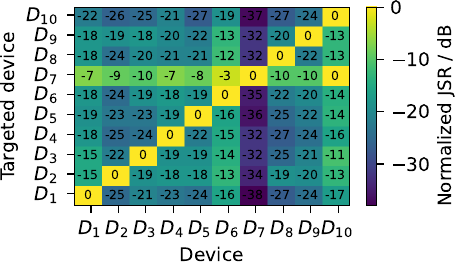}
        \caption{}
        \label{fig:evaluation:selective_jamming:optimization:jsr_norm}
    \end{subfigure}
    \caption{(a) \ac{RSSI} from the jammer as observed for each device. (b) \ac{RSSI} from the \ac{AP}. (c) \ac{JSR} of the \ac{RSSI} from the jammer and the \ac{AP}. (d) Normalization of the \ac{JSR}.}
    \label{fig:evaluation:selective_jamming:optimization}
\end{figure}

As discussed in~\autoref{sec:attack_strategy}, the \ac{JSR} assesses the attacker's success, relating the jamming power at a specific device to the desired signal power from the \ac{AP}. To evaluate the \ac{JSR}, we measure the \ac{RSSI} values from the jammer and the \ac{AP} (both sending \mbox{Wi-Fi} signals with constant transmission power) on each device for each optimized \ac{RIS} configuration. \autoref{fig:evaluation:selective_jamming:optimization:rssi_attacker} and~\autoref{fig:evaluation:selective_jamming:optimization:rssi_accesspoint} present the results, visualizing the effect of the respective \ac{RIS} configuration (per column) as observed on each device (per row). The distinct diagonal entries in~\autoref{fig:evaluation:selective_jamming:optimization:rssi_attacker} show that the attacker succeeds in focusing their signals on the intended devices while achieving rejection towards others, while \autoref{fig:evaluation:selective_jamming:optimization:rssi_accesspoint} depicts each device's signal reception from the \ac{AP}. Here, the \ac{RIS} does not affect the channels  $H^{AP}_{D_i}$ between the \ac{AP} and the devices. Additionally, we observe the effects of distance-dependent path loss, as the devices with the smallest and largest distance to the jammer and \ac{AP} ($D_7$ and $D_6$) experience the highest and lowest signal strengths, respectively. Similarly, $D_7$ receives signals from the \ac{AP} strongest as it is only \SI{1}{m} apart.

Using the \ac{RSSI} measurements from both the jammer and the \ac{AP}, we derive the \acp{JSR} by taking their difference (since \ac{RSSI} values are logarithmic, this follows~\autoref{eq:jsr}). The result is shown in \autoref{fig:evaluation:selective_jamming:optimization:jsr_raw}. The key observation here is that the \ac{JSR} values of the targeted devices on the diagonal entries stand out as desired. 
Furthermore, the variation of the per-target \ac{JSR} indicates that jamming of, \eg, $D_{10}$ is more efficient than jamming $D_{7}$, which has a robust legitimate signal reception. 

To highlight the attack effect on non-target devices, we additionally show the row-wise normalized \ac{JSR} in \autoref{fig:evaluation:selective_jamming:optimization:jsr_norm}. This highlights the different legitimate channel conditions: $D_7$ consistently shows the lowest non-target \ac{JSR} due to its strong signal reception from the \ac{AP}, whereas $D_6$ and $D_{10}$ exhibit higher non-target \acp{JSR} because they receive weaker signals from the \ac{AP}. Finally, the higher relative jamming signal power required for $D_7$ reduces the signal rejection margins towards the other devices such that the attacker fails to exclude~$D_{10}$. Nonetheless, the attacker achieves \ac{JSR} reductions of at least \SI{20}{dB} in more than \SI{50}{\percent} of the cases (\SI{16}{dB} in \SI{90}{\percent} and \SI{24}{dB} in \SI{25}{\percent}). We refer the reader to~Appendix~\ref{sec:attack_evaluation:optimization_effect_jsr} for an evaluation of the normalized~\ac{JSR} during the \ac{RIS} optimization.

\begin{figure}
    \centering
    \includegraphics[width=.70\plotwidth]{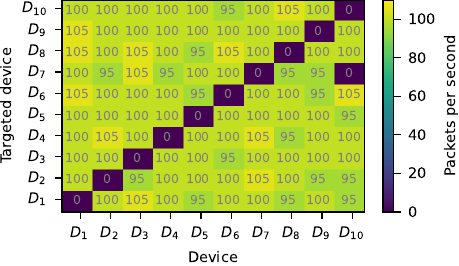}
    \caption{Packet receive rate for each device and for each single-target configuration $D_1$ to $D_{10}$.}
    \label{fig:evaluation:selective_jamming:jamming_result:target}
\end{figure}

Now, while using just enough jamming signal power to disrupt the respective target device, we perform the active jamming attack against each device. Following the same evaluation rationale as before, \autoref{fig:evaluation:selective_jamming:jamming_result:target} shows the packet reception rates of each device while sending the jamming signal with the corresponding \ac{RIS} configuration applied. After demonstrating selective jamming of $D_1$ before, the clearly visible diagonal entries show that the attacker is capable of selectively jamming all considered devices without affecting the other devices. However, we also recognize the effect of insufficient jamming signal rejection, as discussed before. That is, when jamming $D_7$, the attacker also disrupts the packet reception of $D_{10}$, being in line with the previous \ac{JSR} result shown in~\autoref{fig:evaluation:selective_jamming:optimization:jsr_norm}.

\begin{figure}
\centering
    \begin{subfigure}{0.49\columnwidth}
        \centering
        \includegraphics[width=\columnwidth]{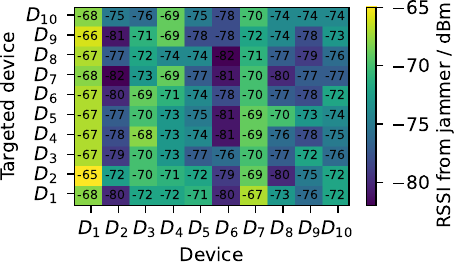}
        \caption{}
        \label{fig:evaluation:selective_jamming:random:rssi_attacker}
    \end{subfigure}
    \begin{subfigure}{0.49\columnwidth}
        \centering
        \includegraphics[width=\columnwidth]{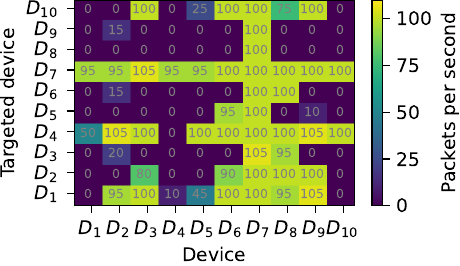}
        \caption{}
        \label{fig:evaluation:selective_jamming:random:rates}
    \end{subfigure}
    \caption{Attack performance with randomized \ac{RIS} configurations. (a) Received \ac{RSSI} from the attacker for each device. (b) Resulting packet reception rates during jamming.}
    \label{fig:evaluation:selective_jamming:random}
\end{figure}

Overall, our results demonstrate that \ac{RIS}-based jamming enables precise physical attack targeting without affecting neighboring devices, even through-the-wall from another room. Please note the reported attack performance clearly stems from the attacker's optimized \ac{RIS}. That is, when using random configurations for the \ac{RIS}, neither the \ac{RSSI} from the attacker nor the packet reception rates under jamming are concentrated on a particular device as evident from~\autoref{fig:evaluation:selective_jamming:random}.

\subsection{Multi-Target Jamming}
\label{sec:attack_evaluation:target_variations}

The previous results highlight the effectiveness of \ac{RIS}-based spatially selective jamming in single-target scenarios. In the following, we extend this scenario and investigate two additional scenarios where the attacker wants to jam not just one, but multiple devices simultaneously. %

\subsubsection{Device Clusters}

\begin{figure}
    \centering
    \begin{subfigure}{0.49\columnwidth}
        \centering
        \includegraphics[width=\columnwidth]{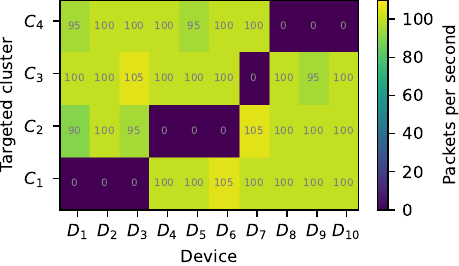}
        \caption{}
        \label{fig:evaluation:selective_jamming:clusters}
    \end{subfigure}
    \begin{subfigure}{0.49\columnwidth}
    \centering
        \includegraphics[width=\columnwidth]{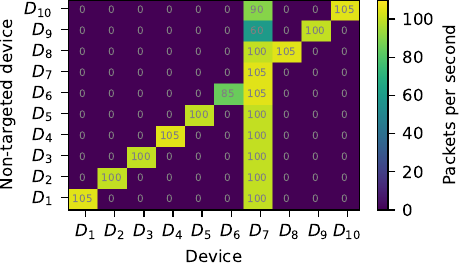}
        \caption{} 
        \label{fig:evaluation:selective_jamming:jamming_result:exclude}
    \end{subfigure}

    \label{fig:evaluation:selective_jamming}
    \caption{(a) Packet reception rates for jamming the four device clusters (see Figure~\ref{fig:experimental_setup}). (b) Packet reception rates for targeting all but one device.}
\end{figure}

In the first multi-device scenario, the attacker seeks to disrupt the device clusters depicted in~\autoref{fig:experimental_setup}. Thus, we repeat the previous experiments but this time specify the devices belonging to the clusters $C_1$ to $C_4$ as target devices during the \ac{RIS} optimization. Subsequently using the resulting four \ac{RIS} configurations for active jamming, \autoref{fig:evaluation:selective_jamming:clusters} shows the packet receive rates of all devices, analogous to~\autoref{fig:evaluation:selective_jamming:jamming_result:target}. Here, we can see that attacker succeeds to disrupt the devices belonging to the respective clusters, while the remaining devices again remain fully operational. Note that the \ac{RIS} configuration to target cluster $C_3$ (comprising only of device $D_7$) this time sufficiently reduces the jamming signal towards device $D_{10}$, preventing the unintended non-target jamming previously observed in~\autoref{fig:evaluation:selective_jamming:jamming_result:target}. That is because the greedy optimization algorithm does not necessarily always converge to the same \ac{RIS} configuration, given that the algorithm is randomly initialized and guided based on noisy measurements. Still, the overall conclusion from this experiment is that the attack approach is extensible to selectively jam even multiple devices simultaneously.

\subsubsection{Single Exclusion}
After jamming multiple devices, we now aim to jam \textit{every} device except one the attacker would like to keep operational.
Building on the previous insight that the attacker can leverage the \ac{RIS} to deliver the jamming signal to multiple devices, we now seek to push this approach even further to jam \textit{every} device except one that the attacker would like to keep operational. Thus, we repeat the \ac{RIS} optimization, but now specify $\mathcal{T} = \mathcal{D} \setminus \{D_0, D_i\}$ where $i \geq 1$. Like before, we then perform active jamming with each optimized \ac{RIS} configuration and plot the packet reception rates of all devices in~\autoref{fig:evaluation:selective_jamming:jamming_result:exclude}. Here, quite opposite to the single-targeting scenario of~\autoref{fig:evaluation:selective_jamming:jamming_result:target}, we can see that the attacker indeed succeeds to disrupt all receivers except $D_7$ while keeping one non-targeted device operational. This experiment confirms that jamming of a broader set of devices is possible.

The inability to jam $D_7$ is due to its strong reception of the legitimate \ac{AP} signal \cf~\autoref{fig:evaluation:selective_jamming:optimization:rssi_accesspoint}. Successful jamming requires the attacker to deliver sufficient signal power to $D_7$ to overshadow the legitimate signal. However, in the present scenario, the attacker splits their jamming power among nine targets, reducing the power efficiency. Thus, we expect the jamming power at each device to be reduced by a factor $1/9\ \widehat{=} -10 \textrm{log}_{10}(9) \approx 9.5~\textrm{dB}$. This matches our observations, as the jamming power arriving at $D_7$ was \SI{-56}{dBm} in the single-target scenario, \cf~\autoref{fig:evaluation:selective_jamming:optimization:rssi_attacker}, while in the current experiment, it was at most \SI{-65}{dBm}. Therefore, the attacker lacks sufficient jamming power to disrupt $D_7$. To address this, the attacker could amplify their jamming signal or prioritize channel maximization towards $D_7$ during \ac{RIS} optimization.

\subsection{Effect of Environmental Variation}
\label{sec:attack_evaluation:environmental_variation}

The attacker relies on the assumption that the optimized \ac{RIS} configuration is valid during the subsequent active jamming. In the following, we investigate the validity of this assumption, evaluating the robustness of the proposed scheme against environmental variation.

\subsubsection{Long-Term Stability}

\begin{figure}
\centering
    \includegraphics[width=0.70\plotwidth]{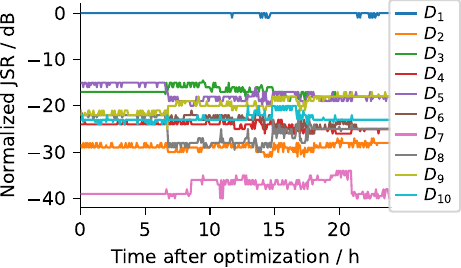}
    \caption{Time stability of optimized \ac{JSR} values over 24~h.} 
    \label{fig:evaluation:longterm_jsr}
\end{figure}

For our previous experiments, the attacker's \ac{RIS} was optimized shortly before obtaining \ac{JSR} and jamming packet reception rate results. However, it is not yet clear whether we can expect the \ac{RIS} optimization to be long-term stable, which would be a desirable property for attack practicality. To investigate this aspect, we perform \ac{RIS} optimization with $\mathcal{T} = \{D_1\}$ and monitor the resulting \ac{JSR} values of each device for a duration of \SI{24}{}~hours. \autoref{fig:evaluation:longterm_jsr} shows the \acp{JSR} normalized to the initial \ac{JSR} measurement of $D_1$ as a time series. The key observation is that the \acp{JSR} remain largely stable with sustained focusing of $D_1$ and rejection of the other devices, showing that the optimized \ac{RIS} configuration stays valid after an entire day has passed. However, we do observe some variation in the \acp{JSR} of the non-target devices, starting~7~hours after the \ac{RIS} optimization at midnight. Since our experiments took place in an ordinary, actively used office environment, we attribute this to office activity, \eg\, individuals walking around.
We investigate the effect of human activity on the attacker's jamming performance more systematically in the next experiment.

\subsubsection{Human Motion}

\begin{figure}
\centering
    \begin{subfigure}{.5\columnwidth}
        \centering
        \includegraphics[width=.8\columnwidth]{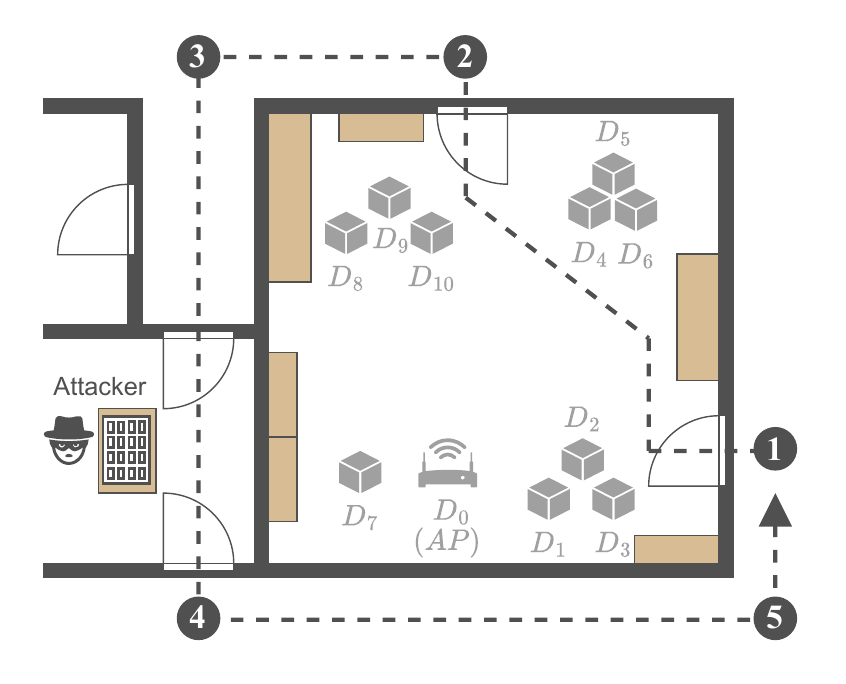}
        \caption{}
        \label{fig:evaluation:environment:path:layout}
    \end{subfigure}%
    \begin{subfigure}{.5\columnwidth}
        \centering
        \includegraphics[width=\columnwidth]{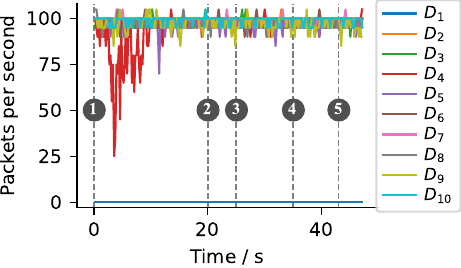}
        \caption{}
        \label{fig:evaluation:environment:path:timeseries}
    \end{subfigure}
    \caption{Effects of human motion in the experimental environment. (a)~Path layout for the motion including checkpoints. (b)~Measured time series for each device.} 
    \label{fig:evaluation:environment:path}
\end{figure}

In this experiment, we again perform single-target jamming with $\mathcal{T} = \{D_1\}$ while an individual walks through the environment, passing by the considered devices and the attacker's \ac{RIS}. At the same time, we record the packet receive rate of each device for \SI{45}{s}. \autoref{fig:evaluation:environment:path:layout} shows the walking path with annotations matching the receive rate time series shown in \autoref{fig:evaluation:environment:path:timeseries}. The first thing to note is that the reception of $D_1$ remains completely suspended as desired, regardless of the individual. Likewise, the reception of the non-targeted devices is mostly unaffected at approx.~$100$~packets per second. Still, we can see that when the individual is within the room, the device $D_4$ is temporarily affected by the jamming as evident from the reduced packet reception. That is, the individual potentially affects the wireless channels $H^{RIS}_{D_4}(c)$ and $H^{AP}_{D_4}$, which caused an increased \ac{JSR}.

\subsubsection{Changes to the Environment}

\begin{figure}
\centering
    \includegraphics[width=.70\plotwidth]{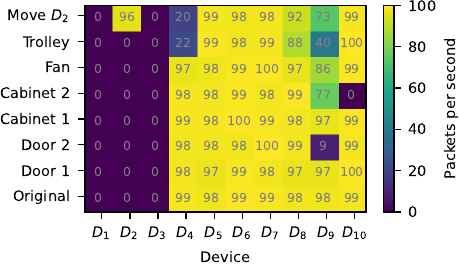}
    \caption{Packet reception rates during jamming with incremental environmental changes.}
    \label{fig:environmental_changes}
\end{figure}

Next, we perform multi-target jamming with $\mathcal{T} = \{D_1, D_2, D_3\}$ (cluster $C_1$) and incrementally change the room where the devices are located. In particular, we open its two doors, introduce additional items (two $60$~$\times$~$60$~$\times$~$43$~cm rolling office cabinets, a pedestal standing fan, and a $60$~$\times$~$60$~cm flat platform trolley), and finally move the device $D_2$ by two centimeters. \autoref{fig:environmental_changes} shows the packet receive rates of all devices for each environmental state. Without changes in the environment (labeled as \lq Original\rq{}), the attacker achieves their goal to jam the target devices while the others remain unaffected. Then, as we introduce more changes to the environment, we can see that especially the non-target devices $D_4$, $D_8$, and $D_9$ become more affected by the jamming, indicating an increased \ac{JSR}. %
As in the previous experiment, jamming of the targeted devices is largely robust against environmental variation. However, when finally moving the targeted device $D_2$, it is no longer disrupted.

\subsubsection{Different Device Positions}
\label{sec:attack_evaluation:simple_jamming:attacker_position}

\begin{figure}%
    \begin{subfigure}{0.49\columnwidth}
        \centering
        \includegraphics[width=\columnwidth]{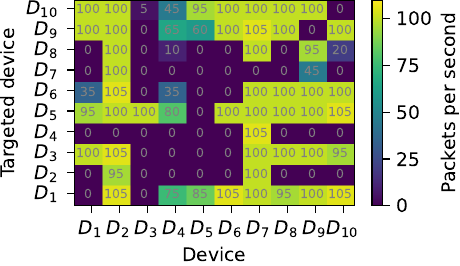}
        \caption{}
        \label{fig:device_positions:attacker_repos}
    \end{subfigure}
    \begin{subfigure}{0.49\columnwidth}
        \centering
        \includegraphics[width=\columnwidth]{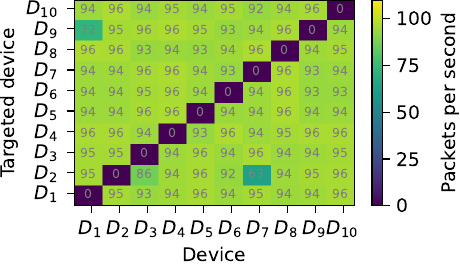}
        \caption{}
        \label{fig:device_positions:attacker_repos_rerun}
    \end{subfigure}
    \caption{Effect of device-repositioning: Selective jamming performance after rearranging the devices in a grid across the room (a) before and (b) after rerunning \ac{RIS} optimization.}
    \label{fig::device_positions}
\end{figure}

In the previous experiment, we have seen that relocating the device $D_2$ caused jamming of that device to become ineffective. This observation underscores the attack's desired dependency on the device location. To further study the effect of relocating devices, we change the device positioning in clusters (see~\autoref{fig:experimental_setup}) to a grid, uniformly distributing the device across the room. Then, we repeat the single-target experiment outlined in~\autoref{sec:attack_evaluation:selective_jamming} with the originally optimized \ac{RIS} configurations. The results are shown in~\autoref{fig:device_positions:attacker_repos}. We now see that the attacker -- using outdated \ac{RIS} configurations -- fails to selectively jam the respective devices. However, after renewing the \ac{RIS} configurations, the attacker is again capable of selectively jamming each device, as can be seen in~\autoref{fig:device_positions:attacker_repos_rerun}. Apart from the clear spatial dependence of the attack, this result also highlights the attacker's ability to adapt to different scenes.

\subsection{Attack Performance in a Wi-Fi Network}
\label{sec:wifi_network_jamming}

Thus far, we studied \ac{RIS}-based selective jamming by means of the \ac{JSR} and packet reception rates. For this, the devices operated in monitor mode while observing packets with a fixed \ac{MCS} setting. This allowed us to evaluate the attack performance independently of the behavior of custom rate adaption algorithms employed in fully operational \mbox{Wi-Fi} networks. Here, the physical-layer transmission speed is not fixed but is adapted to the wireless channel conditions~\cite{xia2013evaluation, lacage2004ieee}. The sender monitors whether transmissions with higher data rates were successful and otherwise reduces the data rate which in turn increases the link robustness. This is controlled via the \ac{MCS} setting~\cite{verges_mcsindex}, denoting a standardized combination of modulation scheme and error coding rate.

In the context of jamming, rate adaption can be viewed as a countermeasure where the victim party adaptively enhances their jamming resiliency. Switching to a lower \ac{MCS} setting allows the victim to cope with reduced link quality or, put differently, with a higher \ac{JSR}. For example, for single-antenna communication over \SI{20}{MHz} bandwidth, \mbox{Wi-Fi} receiver sensitivity improves by approx.~\SI{18}{dB} when switching from MCS~7 to MCS~0~\cite{perahia2008next}, yet reducing data rate by a factor of 10~\cite{verges_mcsindex}.

\begin{figure}
    \centering
    \includegraphics[width=.70\plotwidth]{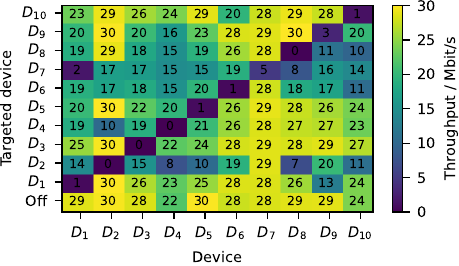}
    \caption{Wi-Fi throughput for each device and for each single-target configuration $D_1$ to $D_{10}$.}
    \label{fig:evaluation:selective_jamming:wifinetwork}
\end{figure}

To assess whether \mbox{Wi-Fi} rate adaption could diminish the attacker's selective jamming success and to further investigate the attack's real-world potential, we repeat the single-target experiment from~\autoref{sec:attack_evaluation:selective_jamming}. However, we measure the data throughput on each device $\{D_1, \dots, D_{10}\}$ in the \mbox{Wi-Fi} network of the \ac{AP}. In particular, we use iperf3 to transfer a \SI{30}{Mbit/s} UDP datastream towards each device while the attacker transmits their jamming signal to disrupt one device. \autoref{fig:evaluation:selective_jamming:wifinetwork} shows the resulting throughput measurements and additionally indicates the throughput without an attack in the first row. Importantly, we again observe clearly distinct diagonal entries, showing that the throughput on the targeted devices is reduced to (nearly) \SI{0}{MBit/s}.

Other than in~\autoref{fig:evaluation:selective_jamming:jamming_result:target}, non-target devices are slightly affected by the jamming. The reason for this is the rate adaption of the targeted device which reacts to the jamming by switching to a more robust \ac{MCS} that can withstand a higher \ac{JSR}. In turn, to achieve a \ac{JSR} that disrupts the target reception, the attacker must expend more jamming power. However, this may exhaust the \ac{JSR} reduction provided by \ac{RIS} optimization, causing a non-target device to be affected and switching to a more robust \ac{MCS}, sacrificing some data rate. Still, the throughput of the non-target devices remains above \SI{22}{Mbit/s} in \SI{50}{\percent} of the cases (\SI{11}{Mbit/s} in \SI{90}{\percent} and \SI{27.5}{Mbit/s} in~\SI{25}{\percent}). 

Consequently, while the attacker does not completely prevent the non-target devices from being affected, the observed throughput degradation is within acceptable limits and can likely be further improved by refining the \ac{RIS} optimization. The results demonstrate the feasibility of the attack even when the victim devices use adaptive rate control, posing a significant threat to real-world \mbox{Wi-Fi} networks.

\subsection{A Detailed Look in the Spatial Domain}
\label{sec:attack_evaluation:detail_spatial}   

Thus far we have investigated \ac{RIS}-based spatially selective jamming of devices distributed across an entire room. Now, we seek to explore the attack on a smaller scale, \ie, when devices are in close proximity.

\subsubsection{Selective Jamming of Devices at Sub-Wavelength Distance}
\label{sec:closeby_iperf}

Previously, we have seen that it is possible to jam single devices although being in close proximity to others within clusters, \cf~\autoref{fig:experimental_setup}. Intuitively, due to the inherent spatial correlation of electromagnetic fields, one would expect that if one device is disrupted by the attacker's jamming, then another very close-by device would likewise be affected. However, interestingly, we found that \ac{RIS}-based spatially selective jamming even works when device antenna separation is deeply in the sub-wavelength region.

\begin{figure}
    \begin{subfigure}{0.5\columnwidth}
        \centering
        \includegraphics[width=\columnwidth]{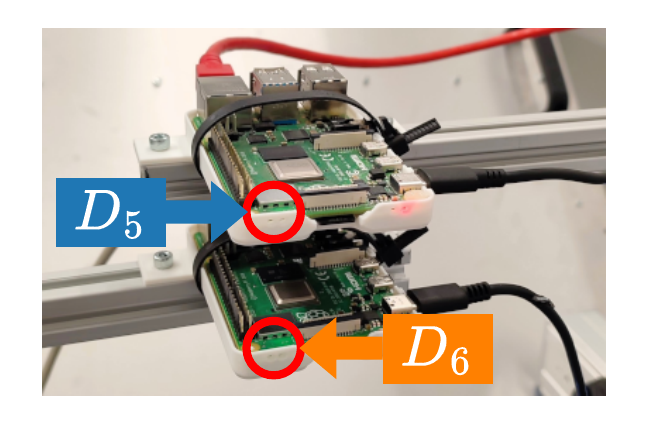}
        \caption{}
                \label{fig:target_setup1:photo}
    \end{subfigure}%
    \begin{subfigure}{0.5\columnwidth}
        \centering
        \includegraphics[width=\columnwidth]{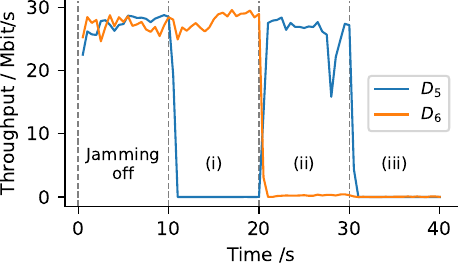}
        \caption{}
        \label{fig:target_setup2:datarates_laptop}
    \end{subfigure}
    \\
    \begin{subfigure}{0.5\columnwidth}
        \centering
        \includegraphics[width=\columnwidth]{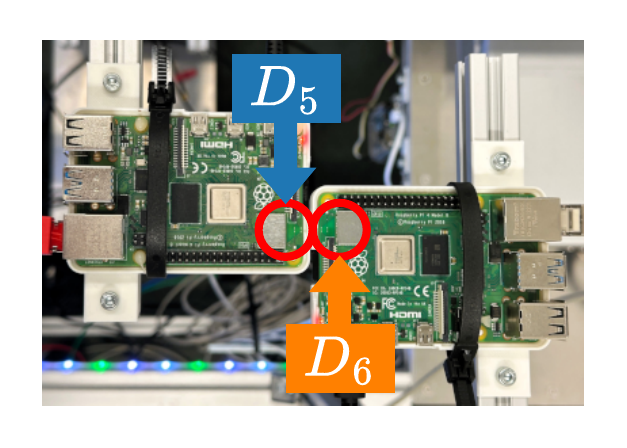}
       
        \caption{}
                \label{fig:target_setup2:photo}
    \end{subfigure}%
    \begin{subfigure}{0.5\columnwidth}
        \centering
        \includegraphics[width=\columnwidth]{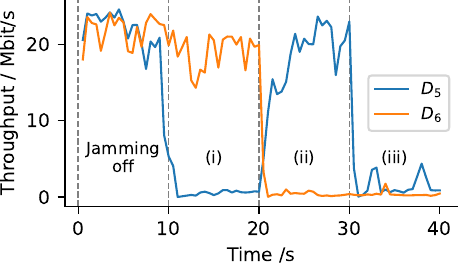}
        \caption{}
        \label{fig:target_setup1:datarates_laptop}
    \end{subfigure}
    \\
    \caption{Real-world spatial jamming attack demonstration against \mbox{Wi-Fi} communication. (a) Parallel aligned antennas. (b) Measured \mbox{Wi-Fi} datarates for parallel antennas. (c) Facing antennas. (d) Measured \mbox{Wi-Fi} datarates for facing antennas.}
    \label{fig:wifi_datarates}
\end{figure}

In our experiment, we consider the devices $D_5$ and $D_6$ (and the \ac{AP} $D_0$) which we place very close to each other in two geometrical configurations. In the first (see~\autoref{fig:target_setup1:photo}), we place the devices directly above each other to minimize their antenna distance while being in the same orientation. In the second (see~\autoref{fig:target_setup2:photo}), we place the devices facing each other to minimize their antenna distance regardless of the orientation, being approx.~\SI{5}{\mm}. For both scenarios, the attacker optimizes their \ac{RIS} to ($i$) target $D_5$, ($ii$) target $D_6$, and ($iii$) target both. Like before, we use iperf3 to measure the \mbox{Wi-Fi} throughput on both devices while the attacker conducts their jamming attack.

The resulting throughput measurements for both device placements are shown as time series in~\autoref{fig:target_setup2:datarates_laptop} and~\autoref{fig:target_setup1:datarates_laptop}. The first thing to note is that without jamming, both devices initially have data rates of around~\SI{25}{Mbit/s}. Then, as the attacker starts to jam with the first \ac{RIS} configuration, the data rate of $D_5$ drops to zero while $D_6$ remains completely unaffected as evident from the unaltered throughput. The attacker then switches to the next \ac{RIS} configuration, alternating the attack target. Now, the throughput of $D_5$ is restored to the level without the attack while the throughput of $D_6$ is close to zero. Finally, as attacker switches to the third \ac{RIS} configuration, the throughput of both devices drops to (nearly) zero. Please note that -- after activating the jamming -- the attacker achieves this result by merely reconfiguring the \ac{RIS} configuration. Importantly, this experiment highlights the attacker's ability to dynamically change the targeted device.

\subsubsection{Further Analysis of Close-By Antennas}
\label{sec:antenna_coupling}

We now investigate the mechanics behind the previous result. First of all, theoretical limits of separating wireless channels in the spatial domain are rooted in the correlation of multipath components at different locations, \ie, the correlations of the $L$ sub-channels via the \ac{RIS} from the attacker antenna towards $D_5$ and $D_6$, $h_l^{D_5}$ and $h_l^{D_6}$. As shown by Clarke~\cite{6779222}, the correlation as a function of the distance~$d$ can be described using the Bessel function of the first kind. Given the attack scenario where one channel is maximized, the smallest distance from the maximized location to a minimum could be approximated as the first zero point of the Bessel function, given by $2.4048\frac{c}{2\pi f} \approx \SI{20.6}{mm}$, where $c$ is the speed of light and $f$ is the signal frequency (\SI{5560}{\MHz} in our experiments). However, this number is considerably higher than the \SI{5}{mm} device separation from~\autoref{fig:target_setup2:photo}. One reason for this is because two antennas in real-world scenarios will seldom exhibit the exact same radiation patterns. This effect is due to differences in relative orientation (such as in~\autoref{fig:target_setup2:photo}) as well as differences in the relative environment, \eg, objects in the antenna nearfield. Notably, the latter also includes the case that antennas get into each other's proximity: So called mutual coupling effects distort the individual antenna radiation patterns and therefore reduce spatial correlation effects~\cite{stjernman2006antenna, derneryd2004signal}. In consequence, the aforementioned effects would allow the attacker, for example, to exploit that one device's antenna might exhibit a high sensitivity in an angular direction where the other does not. 

To assess the influence of the device antennas, we conducted additional experiments using a Keysight~P9372A \ac{VNA} for high-accuracy wireless channel measurements. At the position of $D_8$, we place two antennas of the same type (\lq Antenna~1\rq{} and \lq Antenna~2\rq{}) directly next to each other and measure the wireless channels between them and the attacker's antenna (via the \ac{RIS}). We optimize the \ac{RIS} to maximize the channel towards Antenna~1 and minimize the channel towards Antenna~2 and vice versa. Then, for both cases, we move Antenna~2 in steps of \SI{4}{mm} away from Antenna~1. To ensure accurate and repeatable antenna positioning, we use a 3D-printed positioning fixture. A photo of the setup is shown in~\autoref{fig:antenna_coupling:photo} in Appendix~\ref{appendix:experimental_setup}. 

\autoref{fig::vna_antenne_move_experiment} shows the magnitude channel measurement results for both antennas over the tested displacements of Antenna~2. Here, we can see that the channel of Antenna~2 clearly decorrelates with its displacement. Notably, the measurement is in good agreement with theory, where the signal power is strongly reduced at the zero point of the Bessel function, as indicated by the dashed line~\autoref{fig:vna_move_maximized}. Furthermore, it is also evident that the measurement results of the fixed Antenna~1 are affected by moving Antenna~2. Given the stronger relative impact of small channel variations, this behavior is more pronounced when Antenna~1 is minimized (power increases by approx.~\SI{8}{dB}) than when it is maximized (power reduced by approx.~\SI{0.5}{dB}). Finally, these results confirm that the attacker can take advantage of antenna coupling effects to selectively target devices in close proximity configurations.

\begin{figure}%
    \begin{subfigure}{0.49\columnwidth}
        \centering
        \includegraphics[width=1\columnwidth]{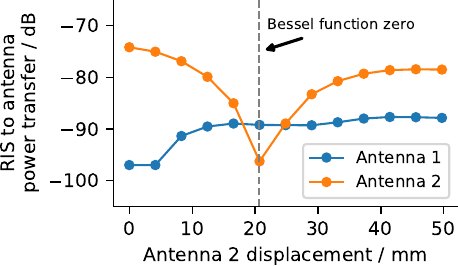}
        \caption{}
        \label{fig:vna_move_maximized}
    \end{subfigure}
    \begin{subfigure}{0.49\columnwidth}
        \centering
        \includegraphics[width=1\columnwidth]{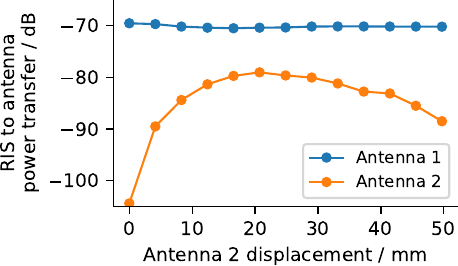}
        \caption{}
        \label{fig:vna_move_minimized}
    \end{subfigure}
    \caption{Channel effects when repositioning an antenna in close proximity of another one. \ac{RIS} optimized with Antenna~2 at \SI{0}{\mm} displacement to (a) minimize and maximize and (b) maximize and minimize the respective channels.}
    \label{fig::vna_antenne_move_experiment}
\end{figure}

\subsubsection{Jamming Impact on Hidden Devices}
\label{sec:attack_evaluation:jamming_hidden_devices}
\begin{figure}
\centering
    \begin{subfigure}{0.49\columnwidth}
        \centering
        \includegraphics[width=\columnwidth]{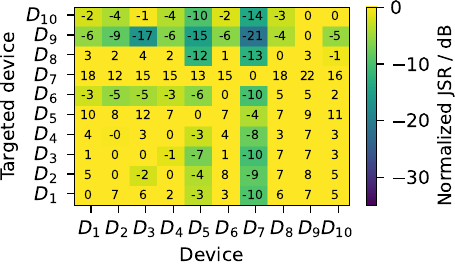}
        \caption{}
        \label{fig:hidden_initial}
    \end{subfigure}
    \begin{subfigure}{0.49\columnwidth}
        \centering
        \includegraphics[width=\columnwidth]{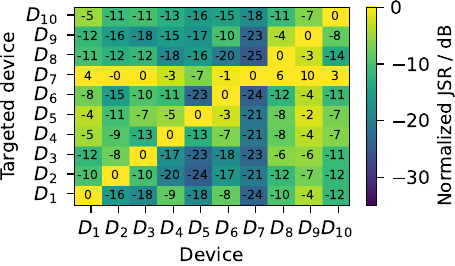}
        \caption{}
        \label{fig:hidden_final}
    \end{subfigure}\\
    
    \caption{Normalized \ac{JSR} for single-target jamming with other devices being hidden before (a) and after (b) optimization.%
    }
    \label{fig:evaluation:hidden_excluded_devices}
\end{figure}

Thus far, we have focused on devices detectable via their wireless transmissions. However, some devices may remain passvie and only act as receivers without transmitting. Next, we examine the effects of attacks on these \textit{hidden devices}, where the attacker cannot optimize their wireless channel.
To align with our system model from~\autoref{sec:system_model}, we define a subset of devices, denoted as $\mathcal{H} \subseteq \mathcal{N}$, which is ignored for the optimization problem in~\autoref{eq:minH}, effectively treating them as hidden. %
We replicated the single-target experiment from~\autoref{sec:evaluation:jsr} but consider $\mathcal{H} = \{D_1, \dots D_{10}\} \setminus \mathcal{T}$, leaving only the \ac{AP} $D_0$ effective within the non-target devices $\mathcal{N}$. Thus, we optimize the \ac{RIS} to maximize the channel of the respective target while minimizing the channel of the \ac{AP} and ignoring all others. Using the resulting \ac{RIS} configurations, we then measure the \ac{JSR} at each device. 
We show the normalized \ac{JSR} with respect to each target device prior to and after the optimization process in~\autoref{fig:evaluation:hidden_excluded_devices}. 
Initially, most hidden devices experience a \ac{JSR} at least as high as the respective target device. Crucially, after the optimization, we observe a \ac{JSR} concentration on the diagonal, indicating that the attacker achieves selective jamming, despite the devices being hidden. The reason for this is that the channel towards the respective target is maximized while the channel towards the hidden devices is not, effectively reducing the jamming interference at the hidden devices. However, the \acp{JSR} at the hidden non-target devices are higher than when they are not hidden, \cf~\autoref{fig:evaluation:selective_jamming:optimization}, which is due to the lack of the additional minimization.

\begin{figure}%
    \centering
    \includegraphics[width=0.75\plotwidth]{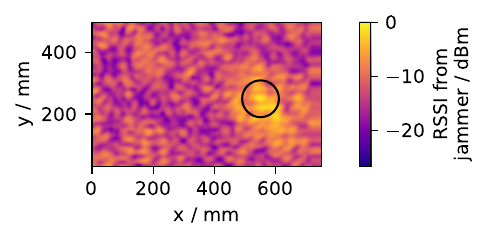}
    \label{fig:attack_scenario:irs_config_heatmap:config_1}
    \caption{Spatial distribution of normalized attacker signal \ac{RSSI} values. During \ac{RIS} optimization, the targeted device was placed within the black circle.}
    \label{fig::robot_heatmap}
\end{figure}

The previous result can be explained by the \ac{RIS} maximization resulting in a focal point around the targeted device, as previously described by  Kaina~\etal~\cite{kainaShapingComplexMicrowave2014}. This is different from classical beamforming, which rather affects an area and not a particular spot. To validate this, we mounted device $D_5$ on a precision dual-axis Cartesian robot, optimized the \ac{RIS} with $\mathcal{T} = \{D_5\}$, and then measured the \ac{RSSI} of the jamming signals at $D_5$ while re-locating the device. \autoref{fig::robot_heatmap} shows the resulting jamming signal distribution, measured in \SI{10}{\mm} steps within an area of size $75$~cm~$\times~50$~cm around the initial device position at $(x,y) = (550,250)$~mm, normalized to the initial position. At positions at least \SI{6}{\cm} away from the initial position (indicated by the black circle), the attacker signal power is at least~$5$~dB and on average~$13$~dB lower. The lesson from this experiment is that one can expect a jamming signal reduction at passive hidden devices, \ie, without explicitly enforcing channel minimization during \ac{RIS} optimization.

\subsection{Further Evaluation of the \ac{RIS}}

\subsubsection{Effect of Surface Size}
\label{sec:attack_evaluation:surface_size}
\begin{figure}
    \begin{subfigure}{0.49\columnwidth}
        \centering
        \includegraphics[width=\columnwidth]{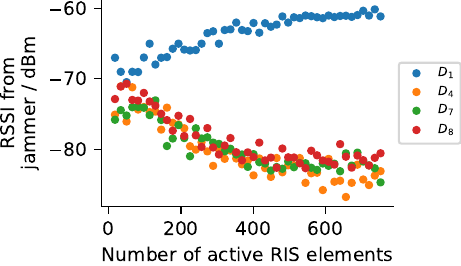}
        \caption{}
        \label{fig:number_ris_elements:k5}
    \end{subfigure}
    \begin{subfigure}{0.49\columnwidth}
        \centering
        \includegraphics[width=\columnwidth]{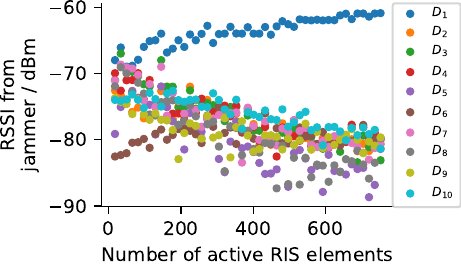}
        \caption{}
        \label{fig:number_ris_elements:k11}
    \end{subfigure}
    \caption{\ac{RIS} optimization results over increasing number of active \ac{RIS} elements for (a)~$K=5$ and (b)~$K=11$ devices.}
    \label{fig:number_ris_elements}
\end{figure}

A relevant factor for the attacker's ability to target and exclude devices is the physical size of the \ac{RIS}~\cite{wuSmartReconfigurableEnvironment2020}, motivating the following experiment. To simulate variations in \ac{RIS} size, we vary the number of active \ac{RIS} elements. Specifically, we perform \ac{RIS} optimization using a randomly selected subset of the total available \ac{RIS} elements, while the other elements are configured random but remain fixed. In this way, we optimize the surface for a single-targeting scenario with $\mathcal{T} = \{D_1\}$ while varying the number of active \ac{RIS} elements from~$16$ to~$768$. We plot the resulting attacker signal strength at each device over the number of active \ac{RIS} elements used for optimization for $K=5$ and $K=11$~total devices in~\autoref{fig:number_ris_elements}.

Here, we see that the attacker fails to separate their \ac{RSSI} values on targeted and non-targeted devices when using less than approx.~100 active elements during the \ac{RIS} optimization. Another observation is that the resulting jammer power levels do not significantly improve beyond approx.~500~elements, potentially opening the door utilizing smaller \acp{RIS}. However, the performance saturation might also be attributed to the particular parametrization of the greedy optimization algorithm, calling for further investigation. Nonetheless, a larger \ac{RIS} improves the attacker's signal control.

Additionally, this experiment also highlights the effect of increasing the number of considered devices $K$. Arguably, as the attacker has to consider more devices, the complexity of the \ac{RIS} channel optimization problem increases. Thus, the optimization algorithm must balance the performance among all devices in the corresponding sets $\mathcal{T}$ and $\mathcal{N}$, potentially sacrificing the optimization quality of one particular device in favor of another.

\subsubsection{Comparison against Directional-Antenna Jamming}
\label{sec:attack_evaluation:comparison}

Alternatively to the \ac{RIS}, the attacker could attempt to use a directional antenna pointed at a targeted device. To investigate whether such an approach could be effective, we perform another jamming experiment where the attacker points an \mbox{elboxRF~TetraAnt~5~19~20~RSLL}~directional antenna with \SI{19}{dBi} gain towards the targeted device $\mathcal{T} = \{D_8\}$. 

\begin{figure}%
    \begin{subfigure}{0.49\columnwidth}
        \centering
        \includegraphics[width=\columnwidth]{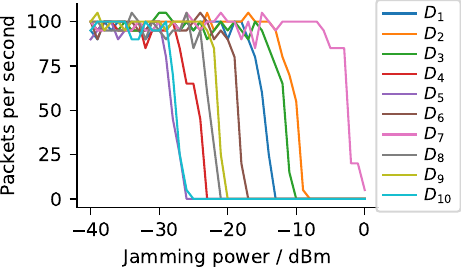}
        \caption{}
        \label{fig:antenna_compare:directional}
    \end{subfigure}
    \begin{subfigure}{0.49\columnwidth}
        \centering
        \includegraphics[width=\columnwidth]{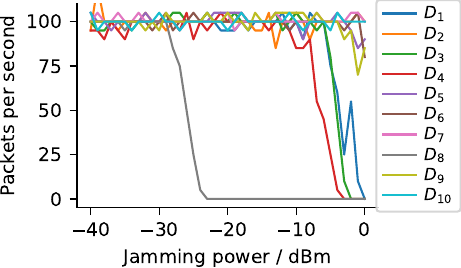}
        \caption{}
        \label{fig:antenna_compare:ris_on_d8}
    \end{subfigure}
    \caption{Comparison between a directional antenna and the RIS: (a) Directional antenna pointed towards $D_8$ and (b) \ac{RIS} jamming, optimized towards with $\mathcal{T} = \{D_8\}$.}
    \label{fig:antenna_compare}
\end{figure}

\autoref{fig:antenna_compare} shows the packet receive rates of all devices over the jamming signal power both for the directional antenna and for the \ac{RIS}. First of all, in both cases the attacker succeeds to disrupt the reception of $D_8$ completely. However, with the directional antenna, the required jamming power is \SI{3}{dB} higher than with the \ac{RIS}. When using the directional antenna, the attacker also jams the non-target devices $D_5$, $D_{10}$, $D_4$, and $D_9$. Furthermore, the jamming power margin until affecting another device ($D_6$) is only \SI{4}{dB}. In clear contrast, with the \ac{RIS}, the attacker succeeds to only jam the device $D_8$ while achieving a jamming power margin of \SI{20}{dB} before another device ($D_4$) is affected.

This result shows that the \ac{RIS} significantly outperforms the directional antenna when considering power efficiency and device selectivity. %
Directional antennas are designed to radiate a single beam under ideal (free-space) conditions, requiring mechanical adjustment while flexibility is limited: Target devices need to be within the beam width and non-target devices must be sufficiently separated. In contrast, the \ac{RIS} allows fully-electronic scene adaption, combining many different propagation paths, allowing focusing and nulling of energy at one or multiple targets devices, regardless of whether they are very close or further apart from each other.

%% file: sections/discussion.tex
\section{Discussion}
\label{sec:discussion}

In this section, we discuss the experimental setup and our results, limitations of the attack, reason about potential countermeasures, and provide directions for future research.

\subsection{Experimental Setup and Results}

\paragraph{Wireless Devices}
We designed the experimental setup to explore the \ac{RIS} for wireless jamming attacks in a realistic scenario, yet made some simplifications to aid experimentation. While the device population comprises of off-the-shelf \mbox{Wi-Fi} devices located in an ordinary office environment, we made the devices regularly ping the \ac{AP} to trigger wireless traffic required for the attacker's \ac{RIS} optimization. In a real-world scenario, the attacker has to contend with the available traffic. However, to tackle potential low transmission rates, the attacker could provoke transmissions by injecting fake packets that devices often respond to with acknowledgements~\cite{abediWiFiSaysHi2020, abediNoncooperativeWifiLocalization2022}. Furthermore, to evaluate the physical-layer mechanism underlying the devised attack scheme while avoiding potential bias due to vendor-specific adaptive device behavior, we relied on monitor mode packet reception for several experiments. To this end, our \mbox{Wi-Fi} throughput measurements demonstrate that the scheme still works when the devices communicate within a fully-fledged \mbox{Wi-Fi} network.

\paragraph{Attacker Setup}
All components for our attacker implementation are either available commercially or open-source, promoting reproducibility of the results. Regarding \mbox{Wi-Fi} signal reception and transmission, the attack can in principle be mounted merely using commodity wireless devices offering monitor mode and packet injection, \eg, using low-level \mbox{Wi-Fi} chipset firmware control~\cite{nexmon:project}. We utilized a dedicated external antenna with the attacker's Raspberry~Pi to illuminate the \ac{RIS}. However, in principle, the attack can be mounted with any antenna, provided the signal reaches the \ac{RIS}. 

The time for each \ac{RIS} optimization step is governed by the time it takes until a signal from each device has been received and therefore depends on the packet transmission rate of \mbox{Wi-Fi} devives which often is higher than $100$~packets per second~\cite{zhuTuAlexaWhen2020}. In our setup, optimization of the \ac{RIS} with \SI{10000}{}~steps takes approx.~5~minutes to finish. When device positions remain static, the attacker is not limited by optimization time, yet to operate in more dynamic environment, optimization speed becomes more relevant. As discussed in Appendix~\ref{sec:attack_evaluation:optimization_effect_jsr}, the number of optimization steps can be substantially reduced without heavily sacrificing performance. 

For the sake of experimental simplicity, we used device \ac{MAC} addresses to distinguish signals from different devices. A fully payload-agnostic physical-layer attacker would have to rely on physical-layer measures such as radio transmitter fingerprints~\cite{jooHoldDoorFingerprinting2020, smailesWatchThisSpace2023, soltanieh2020review, danev2012towards} for this.

While \acp{RIS} are already commercially marketed~\cite{greenerwave_website}, there is no broad consumer-level availability. However, the \ac{RIS} we used is based on the well-documented open-source design by Heinrichs~\etal~\cite{heinrichsOpenSourceReconfigurable2023} who also provide detailed manufacturing information. We estimate that the parts to manufacture the 768-element \ac{RIS} we used can be purchased for approximately~\EUR{750}. The way we leverage the \ac{RIS} resembles an electronically tunable \textit{reflectarray antenna} which was thus far exclusive to high-profile applications such as air surveillance radar~\cite{rosado2021design} or satellites~\cite{karimipourShapingElectromagneticWaves2019}. 
With the advent of \acp{RIS}, such technology now becomes accessible even for individuals, extending the toolkit for advanced physical-layer attacks and necessitating a re-evaluation of attacker capabilities.

\subsection{Countermeasures}

Wireless jamming itself cannot be prevented due to the broadcast nature of radio wave propagation. Instead, jamming-resistant modulation schemes such as spread-spectrum techniques can be used to enhance robustness against jamming. However, changing the physical-layer modulation scheme is not an option for conventional standard-compliant wireless communication systems such as \mbox{Wi-Fi}. Therefore, in the context of this work, we discuss potential countermeasures geared towards hampering the attacker's \ac{RIS} optimization.

\paragraph{MAC Address Randomization}
As a raise-the-bar countermeasure, dynamic randomization of MAC~addresses would make it more difficult for the attacker to associate received wireless signals with specific devices. However, the adversary could utilize payload-independent physical-layer properties such as radio frequency fingerprints~\cite{givehchianEvaluatingPhysicallayerBle2022, smailesWatchThisSpace2023, jooHoldDoorFingerprinting2020} to distinguish radio signals from different devices.

\paragraph{Randomizing Transmit Power}
The attacker's \ac{RIS} optimization during attack preparation relies on \ac{RSSI} values obtained from eavesdropped wireless signals. Thus, as an ad hoc countermeasure, a victim party could randomize their transmit power to hamper the \ac{RIS} optimization. However, this would imply a reduced wireless communication quality of service. However, the attacker could also observe fine-grained \ac{CSI} values which are not affected much by moderate changes of signal power.%

\paragraph{Randomized Transmit Beamforming}
Devices with multiple antennas could employ randomized transmit beamforming during their wireless communication. This would yield randomization of the channel $H^{D_i}_{RIS}(c)$ towards the eavesdropping attacker who uses $H^{D_i}_{RIS}(c)$ to optimize their \ac{RIS}. In consequence, the attacker would be unable to assess whether channel changes stem from the \ac{RIS} or the victim's transmit beamforming, hampering \ac{RIS} optimization.

\paragraph{Avoiding Channel Reciprocity}
A key mechanism underlying the attack is channel reciprocity, allowing the attacker to passively adapt their jamming channel before launching the active attack. Thus, to hamper attack preparation, reciprocal channels should be avoided, \eg, by using sufficiently separated frequencies or antennas for reception and transmission.

\paragraph{Attack detection}
Since the jamming signal is not fully suppressed at non-target locations, \cf our evaluation of attack effects towards hidden devices in Appendix~\ref{sec:attack_evaluation:jamming_hidden_devices}, passive wireless receivers can be used to detect the jamming signals, permanently monitoring the wireless environment, \eg, to raise an alarm upon detecting malicious activity.

\subsection{Limitations}
We have shown that the \ac{RIS} enables precise spatial control for targeted wireless jamming. However, as certain preconditions must be met for this, the attack also is subject to limitations. First of all, like in every other jamming attack, sufficient jamming signal power must be delivered to disturb the victim receiver. While our scheme offers fine-grained spatial jamming control at considered device locations, the jamming effect does not completely vanish at other locations. 

To passively optimize the jamming channel towards the victim device, the attacker relies on passive eavesdropping and a reciprocal wireless channel. Thus, the attack does not work with wireless systems that rely on non-reciprocal wireless channels, \eg, when transmission and reception employ different signal frequencies or antennas. To overcome this, the attacker could perform active jamming while observing whether the victim's throughput is disrupted to indirectly infer the quality of the jamming channels $H^{RIS}_{D_i}(c)$.

Our attack is geared towards bidirectional wireless communication devices that not only receive but also transmit RF signals which is crucial for the attacker to optimize their jamming channel. Therefore, completely passive radio receivers in unidirectional systems, \eg, media broadcasting or satellite navigation, cannot be targeted as the attacker has no means of optimizing their \ac{RIS}.

\subsection{Future Work}
In this work, we have studied selective targeting of individual receivers on the physical layer. Opposite to that, it would also be possible to target individual transmitters by employing transmitter-reactive jamming. More work is needed to assess the feasibility of such an approach and how it compares to ours. Moreover, the combination of spatial and time-varying jamming techniques provides an interesting opportunity to realize spatio-temporal jamming, \eg, to enhance stealthiness. For instance, time-varying modulation of the \ac{RIS} during jamming could be used for effective multi-device targeting, changing between a set of single-target \ac{RIS} configurations.
Although our attacker implementation already yields satisfactory results, we believe the hardware setup can be further improved, \eg, realizing the attack with a single-chip wireless transceiver or using different \ac{RIS} designs, possibly promoting hardware miniaturization. We employed a greedy algorithm from the literature to optimize the~\ac{RIS}. We believe that this process can be further improved, \eg, by studying alternative algorithms, including machine learning-based approaches that might be capable of one-shot synthesis after an initial training. Using more fine-grained \ac{CSI} channel measurements would likely aid faster and more accurate convergence while enabling spatio-spectral control of jamming signals.

%% file: sections/related_work.tex
\section{Related Work}
\label{sec:related_work}

\paragraph{Differentiation from Previous Work}
Previous research has investigated the adversarial use of \acp{RIS} for jamming, yet with a clear focus on \textit{passive} attacks. In particular, there are two main approaches: The first, proposed by Lyu~\etal~\cite{lyuIRSBasedWirelessJamming2020}, is to use the \ac{RIS} to reflect legitimate signals in way that a cancellation signal is formed at the targeted receiver, interfering destructively and thus reducing the received signal power. The second approach, first proposed by Staat~\etal~\cite{staatMirrorMirrorWall2022}, leverages the \ac{RIS} to create fast environmental variation which disturbs a targeted \mbox{Wi-Fi} receiver. Both approaches manipulate legitimate signals which is the key difference compared to our work: We \textit{actively} transmit a jamming signal while using the \ac{RIS} for precise attack targeting.

Karlsson~\etal~\cite{karlsson2014massive, karlsson2017jamming} proposed to exploit channel reciprocity of \ac{TDD} communication systems for jamming attacks. However, different from our work, their goal was to enhance the attacker's power efficiency and not selective jamming. Crucially, they consider an attacker employing a massive \ac{MIMO} radio instead of a single-antenna radio in conjunction with an \ac{RIS} as we do.

\paragraph{Adversarial \ac{RIS} Applications}
Apart from jamming, the \ac{RIS} can be used adversarially to, \eg, manipulate radar sensing, as shown by Vennam~\etal~\cite{reddyvennamMmSpoofResilientSpoofing2023} and Chen~\etal~\cite{chenMetaWaveAttackingMmWave2023}. Zhu~\etal~\cite{zhouRIStealthPracticalCovert2023} have shown that the \ac{RIS} allows attackers to evade wireless sensing-based physical intrusion detection. Other works consider the \ac{RIS} to facilitate eavesdropping, \eg, Chen~\etal~\cite{chenWavefrontManipulationAttack2023}, Chen and Ghasempour~\cite{chenMaliciousMmWaveReconfigurable2022}, and Shaikhanov~\etal~\cite{shaikhanovMetasurfaceintheMiddleAttackTheory2022}. Finally, Li~\etal~\cite{liRISJammingBreakingKey2023} have shown \ac{RIS}-based jamming of wireless key generation.

\paragraph{Jamming Attacks}

An early study on the threat of jamming in wireless communication networks is the work of Xu~\etal~\cite{xuFeasibilityLaunchingDetecting2005}, covering several attack strategies, including constant random signal jamming, deceptive jamming based on packets with valid encoding, time-pulsed jamming, and reactive jamming. These types are also covered in various survey and overview works on attacks and defenses by, \eg, Mpitziopoulos~\etal~\cite{mpitziopoulosSurveyJammingAttacks2009}, Grover~\etal~\cite{groverJammingAntijammingTechniques2014}, Poisel~\cite{poiselModernCommunicationsJamming2011}, and Lichtman~\etal~\cite{lichtmanCommunicationsJammingTaxonomy2016}. Some of these recognize the attacker's antenna characteristics as a degree of freedom or make distinctions between omnidirectional and directional antennas. However, a concept like our scene-adaptive spatially selective jamming is not mentioned. Proano and Lazos~\cite{proanoSelectiveJammingAttacks2010} describe time-domain selective wireless jamming based on real-time packet classification for reactive jamming. Pursuing the same goal, Aras~\etal~\cite{arasSelectiveJammingLoRaWAN2017} describe a packet classification method for LoRaWAN. Reactive jamming has been implemented on smartphones~\cite{schulzMassiveReactiveSmartphoneBased2017} and software-defined radios~\cite{wilhelm2011reactive}, yet can be counteracted using hiding methods as outlined by Proano~\etal~\cite{proano2011packet}. Apart from the general threat of jamming, the literature also presents threat analyses for recent cellular systems such as 4G~\cite{girkeResilient5GLessons2019} and 5G~\cite{arjouneSmartJammingAttacks2020}, \eg, discussing the impact of disrupting certain control channels.

A different line of work addresses the detection of jamming attacks~\cite{strasserDetectionReactiveJamming2010, chiangCrossLayerJammingDetection2011, lyaminRealTimeDetectionDenialofService2014}, more recently also including machine-learning based methods~\cite{testiMachineLearningBasedJamming2023, pawlakMachineLearningApproach2021}. Other works examine friendly jamming, where the goal is to disrupt potential adversaries, \eg, to achieve confidentiality~\cite{wenboshenAllyFriendlyJamming2013, kimCarvingSecureWifi2012}. However, Tippenhauer~\etal~\cite{tippenhauerLimitationsFriendlyJamming2013} and~Robyns~\etal~\cite{robynsPHYlayerSecurityNo2017} have shown that such schemes can be circumvented.

%% file: sections/conclusion.tex
\section{Conclusion}
\label{sec:conclusion}
In this paper, we investigated the merits of the \ac{RIS} technology for active wireless jamming attacks. In particular, we have shown that the \ac{RIS} enables precise physical-layer attack targeting in the spatial domain, enabling protocol level-agnostic selective jamming. For this, the attacker first determines an \ac{RIS} configuration by eavesdropping wireless traffic from the victim devices. Then, the attacker uses the \ac{RIS} to transmit a jamming signal that disrupts the wireless communication of targeted devices while leaving other devices operational. We have demonstrated the effectiveness of the attack under real-world conditions with extensive experimentation using commodity \mbox{Wi-Fi} devices and an open-source \ac{RIS}. Notably, we found that that it is possible to differentiate between devices that are located only millimeters apart from each other. Overall, our work underscores the threat of wireless jamming attacks and recognizes the adversarial potential of \acp{RIS} to enhance the landscape of wireless physical-layer attacks.

\section*{Acknowledgements}
We thank Simon Tewes, Markus Heinrichs, and Rainer Kronberger for providing the \ac{RIS} prototypes and Harald Elders-Boll for discussions. We thank the anonymous reviewers from both the initial and current versions of this paper for their valuable feedback. This work was supported by the Deutsche Forschungsgemeinschaft (DFG, German Research Foundation) under Germany's Excellence Strategy - EXC 2092 CASA - 390781972, RWTÜV Foundation (project number: S0189/10037/2021) and German Federal Office for Information Security (FKZ: Pentest-5GSec - 01MO23025B).

%% file: sections/appendix.tex
\appendix
\section{Appendix}

\subsection{Experimental Setup}
\label{appendix:experimental_setup}

\begin{figure}[!ht]
    \centering
    \includegraphics[width=0.5\columnwidth]{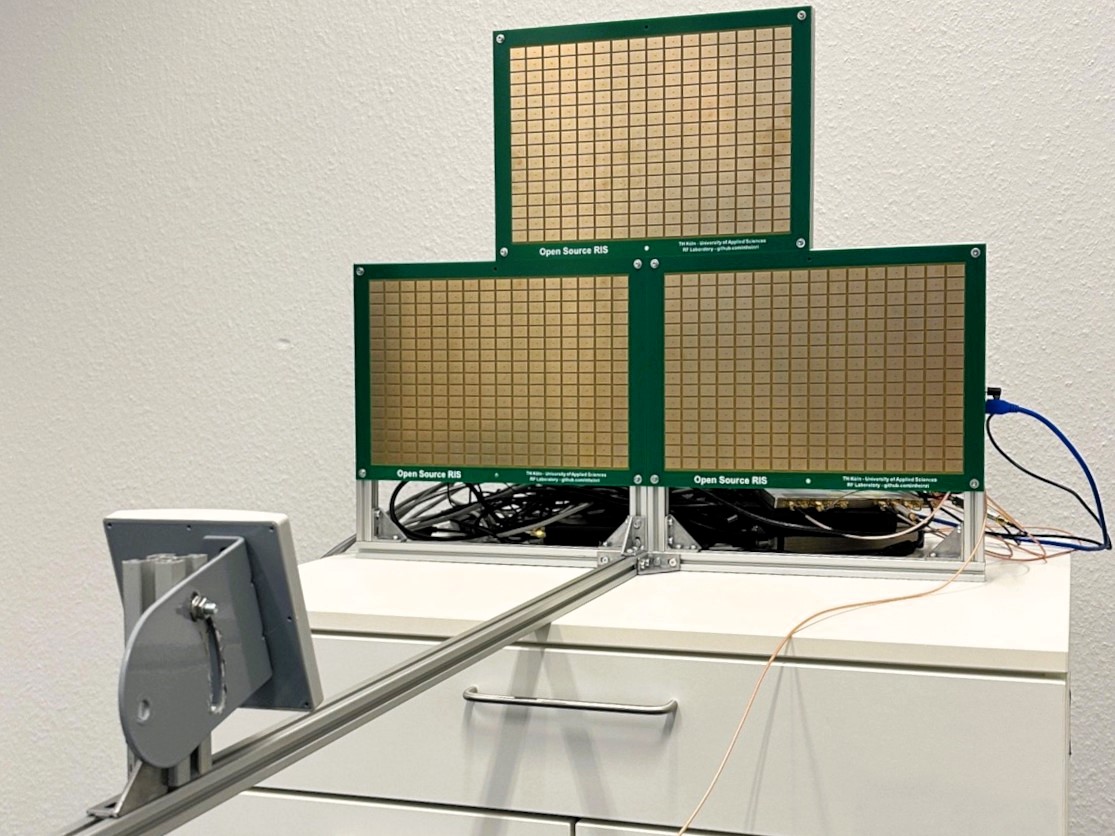}
    \caption{Attacker setup used for experimental evaluation.}
    \label{fig:jamming_setup:photos:attacker}
    \label{fig:jamming_setup:photos}
\end{figure}

Complementing the description in~\autoref{sec:overview:experimental_setup}, we show a photo of the experimental setup in \autoref{fig:jamming_setup:photos}, comprising of the \ac{RIS} that is illuminated by the attacker's antenna.

\subsection{Close-By Antennas}
\label{appendix:close_by_vna}

We here report additional results from the experiment described in~\autoref{sec:antenna_coupling}.

\begin{figure}[!ht]
\centering
    
    \begin{subfigure}{0.45\columnwidth}
        \centering
        \includegraphics[width=\columnwidth]{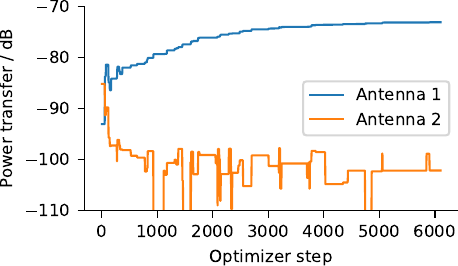}
        \caption{}
        \label{fig:antenna_coupling:optimizer_progress}
    \end{subfigure}
    \begin{subfigure}{0.45\columnwidth}
        \centering
        \includegraphics[width=\columnwidth]{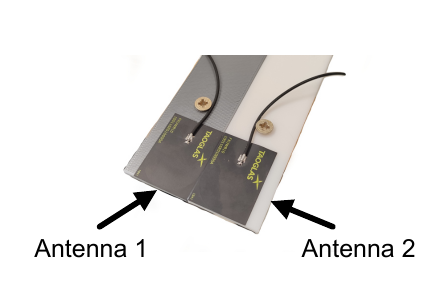}
        \caption{}
    \end{subfigure}\\

    \begin{subfigure}{0.45\columnwidth}
        \centering
        \includegraphics[width=\columnwidth]{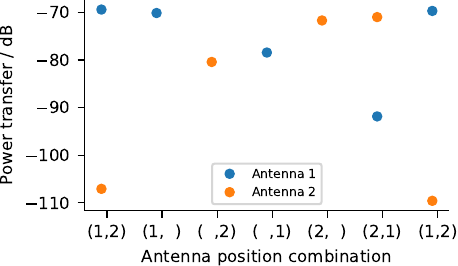}
        \caption{}
        \label{fig:antenna_coupling:antenna_removal}
    \end{subfigure}
    \begin{subfigure}{0.45\columnwidth}
        \centering
        \includegraphics[width=\columnwidth]{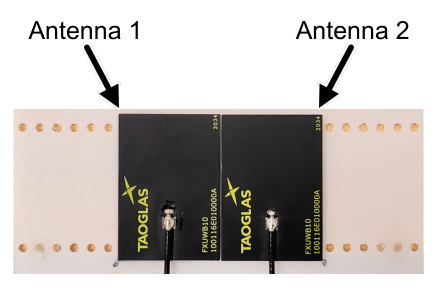}
        \caption{}
        \label{fig:antenna_coupling:photo}
    \end{subfigure}\\
    \caption{Optimization progress (a) of antenna arrangement in (b) using \ac{VNA} measurements. (c) Illustrates effects of removing and reordering antennas, with (1, 2) indicating the arrangement in (d).} 
    \label{fig:antenna_coupling}
\end{figure}

\autoref{fig:antenna_coupling:optimizer_progress} shows the \ac{RIS} optimization process to achieve maximization of the attacker's channel towards Antenna 1 while minimizing the channel towards Antenna 2. The antennas are placed directly next to each other as shown in~\autoref{fig:antenna_coupling:optimizer_progress} and face the attacker's \ac{RIS}. In this experiment, we used a \ac{VNA} to gather channel measurements, confirming the observations previously made with \mbox{Wi-Fi} devices.

In the next experiment, we study the effect of antenna positioning and removal. For this, we utilize a 3D-printed positioning fixture as shown in~\autoref{fig:antenna_coupling:photo}. In \autoref{fig:antenna_coupling:antenna_removal}, we show the channel measurements for various combinations of antenna locations. On the x-axis, we indicate which antenna is located at which position in~\autoref{fig:antenna_coupling:photo}. The first value in parentheses corresponds to the left position, the second value to the right position. After the initial \ac{RIS} optimization, the power transfer towards the two antennas differs by more than \SI{30}{dB}. Then, when removing Antenna~2, the maximized Antenna~1 is barely affected. In contrast, when removing Antenna 2, the initial \SI{-107}{dB} minimization of Antenna~2 deteriorates significantly to around \SI{-80}{dB}. Similar effects are observed when exchanging Antenna~1 and Antenna~2. However, with both antennas present, yet swapped, the initial performance is not matched, indicating slight deviations and imperfections regarding equal antenna positioning. However, when finally repeating the initial measurement, the performance again matches the initial values. This experiment clearly highlights the effects of mutual antenna coupling on the attacker's ability to separate the antenna channels despite close proximity.

\subsection{Behavior of the \ac{RIS} Optimization}
\label{sec:attack_evaluation:optimization_effect_jsr}

\begin{figure}
    \begin{subfigure}{0.49\columnwidth}
        \centering
        \includegraphics[width=\columnwidth]{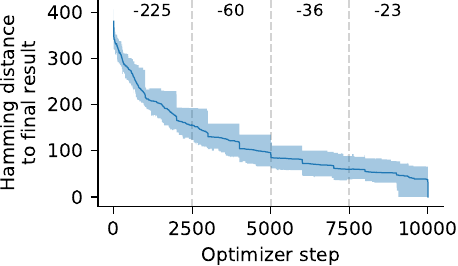}
        \caption{}
        \label{fig:attack_evaluation:further_evaluation:optimization_effect:jsr_evolution:config_distance}
    \end{subfigure}%
    \begin{subfigure}{0.50\columnwidth}
        \centering
        \includegraphics[width=\columnwidth]{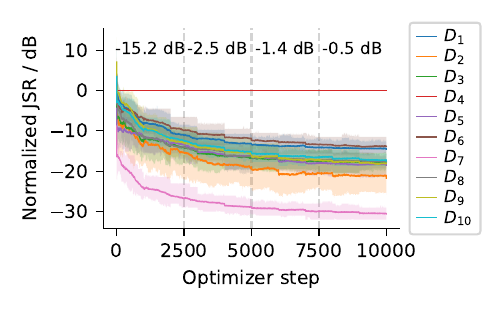}
        \caption{}
        \label{fig:attack_evaluation:further_evaluation:optimization_effect:jsr_evolution:normalized}
    \end{subfigure}%
    \caption{Optimization results from the \ac{RIS} over $50$ runs targeting $D_4$: (a) Hamming distances to the final configuration and (b) measured \ac{JSR} per step. Annotations indicate changes in average Hamming distance and \ac{JSR} over $2500$ steps.}
    \label{fig:attack_evaluation:further_evaluation:optimization_effect:jsr_evolution}
\end{figure}

Throughout this work, we conducted numerous experiments involving optimization of the \ac{RIS} configuration. As outlined in~\autoref{sec:overview:experimental_setup:ris_optimization}, we utilize a greedy heuristic from the literature~\cite{tewes2022full}.
To characterize the consistency and course of the optimization, we performed the single-target experiment from~\autoref{sec:evaluation:jsr} with $\mathcal{T} = \{D_4\}$, repeated $50$~times. For each run and algorithm step, we stored the current best \ac{RIS} configuration and measured the resulting \ac{JSR} at the devices.

Evaluating the optimization convergence speed, we treat \ac{RIS} configurations as 768-bit sequences and calculate the Hamming distances of the final optimization results to previous ones of the optimization progress. The average and 5th and 95th percentiles across repetitions over the optimization progress are shown in~\autoref{fig:attack_evaluation:further_evaluation:optimization_effect:jsr_evolution:config_distance}.  
Due to the random algorithm initialization, we first observe an average difference of~$382$ elements, close to the ideal expected value of~$384$. At the beginning, the \ac{RIS} configuration quickly evolves towards the final result, as evident by the steep reduction of the Hamming distance by $225$ after $2500$~steps. After 4573~algorithm steps, the average Hamming distance to the final optimization result is below $100$~elements. From the percentiles of the distribution, we see that this behavior is largely consistent across different instantiations of the algorithm, regardless of the random initialization. Please note that the staircase pattern stems from periodic re-evaluation of all $B$ \ac{RIS} configuration candidates every $1000$ steps.

In~\autoref{fig:attack_evaluation:further_evaluation:optimization_effect:jsr_evolution:normalized}, we present the corresponding normalized \ac{JSR} during the optimizer progress, again with the 5th and 95th percentiles across algorithm repetitions. Here, it becomes evident that similar \ac{JSR} performance is achieved, regardless of the random algorithm initialization and the inherently noisy \ac{RSSI} measurements. 
In the plot, we annotate the average \ac{JSR} reductions after $2500$ steps, showing only marginal improvement of $1.9$~dB during the last~$5000$~steps. Thus, we conclude that terminating the optimization early is possible without significantly sacrificing \ac{JSR} performance, potentially allowing quicker adaption in dynamic environments.%

\subsection{Cross-Cluster Multi-Target Jamming}
\label{appendix:multi_target_cross_cluster}
\begin{figure}
    \centering
    \includegraphics[width=0.8\plotwidth]{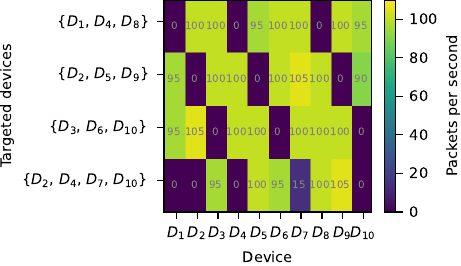}
    \caption{Additional results highlighting the ability to target multiple devices in different clusters.}
    \label{fig:cross_cluster_multi_target_jamming}
\end{figure}

Complementing the results from~\autoref{sec:attack_evaluation:target_variations}, we now demonstrate that it is possible to target multiple devices even when these do not belong to the same device clusters. In particular, we conducted a multi-target jamming experiment where we subsequently targeted the devices $\{D_1, D_4, D_8\}$, $\{D_2, D_5, D_9\}$, $\{D_3, D_6, D_{10}\}$, and $\{D_2, D_4, D_7, D_{10}\}$. The resulting packet rates are illustrated in~\autoref{fig:cross_cluster_multi_target_jamming}, showing successful disruption of one device per cluster. When we include the device $D_7$ to be targeted, which is closest to the access point and hence demands most jamming signal power to be disrupted, we observed that the $D_1$ is unintentionally jammed as well.

